\documentclass[%
 reprint,
 superscriptaddress,
 nofootinbib,
 amsmath,amssymb,
 aps,
 prx,
 floatfix,
 showkeys,
]{revtex4-2}

\usepackage{graphicx}
\usepackage{dcolumn}
\usepackage{bm}

\usepackage[per-mode=symbol]{siunitx}
\usepackage[hidelinks]{hyperref}
\usepackage[capitalize]{cleveref}
\usepackage{braket}
\usepackage{dsfont}
\usepackage{tabularx}
\usepackage{booktabs}

\newtheorem{definition}{Definition}

\makeatletter 

\newcommand{\ketbra}{\@ifstar{\ketbra@star}{\ketbra@nostar}}
\newcommand{\Ketbra}{\@ifstar{\Ketbra@star}{\Ketbra@nostar}}
\newcommand{\ketbra@star}[1]{\ket{#1}\hspace{-1mm}\bra{#1}}
\newcommand{\Ketbra@star}[1]{\Ket{#1}\hspace{-1mm}\Bra{#1}}
\newcommand{\ketbra@nostar}[2]{\ket{#1}\hspace{-1mm}\bra{#2}}
\newcommand{\Ketbra@nostar}[2]{\Ket{#1}\hspace{-1mm}\Bra{#2}}

\makeatother

\newcommand{\genmatrix}[1]{\textrm{#1}}
\newcommand{\system}{\mathsf}  
\newcommand{\genset}{\mathcal}

\DeclareMathOperator{\trace}{Tr}
\newcommand{\ptrace}[1]{\trace_{#1}}

\newcommand{\transpose}{\mathsf{T}}
\newcommand{\adjoint}{\dagger}
\DeclareMathOperator{\diagonal}{diag}

\newcommand{\kron}{\otimes_{\textrm{K}}}

\newcommand{\dvec}{\vec{\Lambda}}
\newcommand{\transmat}{\textrm{T}}  
\newcommand{\bbasis}{\genset{B}}
\newcommand{\qoperation}{\mathcal{E}}

\newcommand{\identity}{\mathds{I}}

\newcommand{\paulix}{{X}}

\newcommand{\pauliz}{{Z}}

\newcommand{\widebar}[2][1.5]{{}\mkern#1mu\overline{\mkern-#1mu#2\mkern-#1mu}\mkern#1mu} 

\newcommand{\logicalw}[1]{{#1}_\text{L}}
\newcommand{\logicalzero}{\logicalw{0}}
\newcommand{\logicalplus}{\logicalw{+}}
\newcommand{\logicalbell}{\logicalw{\Phi}}
\newcommand{\physicalbell}[1]{\Phi}
\newcommand{\codespace}{\genset{W}}

\newcommand{\zbasis}{\genset{Z}}
\newcommand{\xbasis}{\genset{X}}

\newcommand{\bellbasis}{\bbasis_{\text{Bell}}}
\newcommand{\basiszero}{\bbasis_{\text{zero}}}
\newcommand{\basisplus}{\bbasis_{\text{plus}}}
\newcommand{\basisent}{\bbasis_{\text{enc}}}

\newcommand{\totaldist}{L}
\newcommand{\hopdist}{L_0}
\newcommand{\numrepeaters}{N_r}
\newcommand{\repeater}[1]{\system{R}_{#1}}
\newcommand{\numchannels}{{N_{\text{mux}}}}

\newcommand{\initfid}{F_{\text{init}}}
\newcommand{\coherencetime}{T_\text{coh}}
\newcommand{\probgateerr}{\beta}
\newcommand{\probmeaserr}{\delta}

\newcommand{\probcou}{p_\text{cou}}
\newcommand{\detefficiency}{\eta_D}

\newcommand{\chatten}{\alpha_\text{ch}}
\newcommand{\lightspeed}{c}

\newcommand{\secretfrac}{r_{\infty}}
\newcommand{\keyrate}{K}
\newcommand{\distribrate}{R_{\text{dist}}}

\newcommand{\labelencoding}{\text{ENC}}
\newcommand{\labelmemory}{\text{MEM}}
\newcommand{\labelswapping}{\text{SWP}}

\newcommand{\initstate}{\rho^{\text{init}}}
\newcommand{\shapezero}{\dvec_{\ket{\logicalzero}}}
\newcommand{\shapeplus}{\dvec_{\ket{\logicalplus}}}
\newcommand{\shapepair}[1]{\dvec_{\text{g}_{#1}}}

\newcommand{\probgen}{p_\text{gen}}

\newcommand{\avebatches}[1]{\hat{G}_{#1}}

\newcommand{\tgensuccess}[1]{\hat{\tau}_{\text{g}_{#1}}}
\newcommand{\thop}{{\tau}_\text{hop}}
\newcommand{\timeinit}{\tau_{\text{init}}}
\newcommand{\timeattempt}{\tau_{\text{att}}}

\newcommand{\probhop}[1]{p_{\text{h}_{#1}}}
\newcommand{\probendtoend}{p_\text{end}}

\newcommand{\refresult}{r_{\text{ref}}}
\newcommand{\povm}{{P}}

\newcommand{\qchandephase}[1]{\qoperation^{\text{mpf}({#1})}}
\newcommand{\qchanmemdepol}[1]{\qoperation^{\text{mdp}({#1})}}

\newcommand{\shapelogicalplus}{\dvec_{\ket{\logicalplus}}}
\newcommand{\shapelogicalzero}{\dvec_{\ket{\logicalzero}}}

\newcommand{\shapeafteriswaps}[1]{\dvec_{\text{h}_{#1}}}
\newcommand{\shapeafterencoding}{\dvec_{\ket{\logicalbell}}}
\newcommand{\shapeendtoend}{\dvec_{\text{end}}}

\newcommand{\probdecode}[1]{p_{\text{dcd}|{#1}}}

\newcommand{\erate}[2]{e_{{#2}|{#1}}}
\newcommand{\vecerate}[1]{\vec{e}_{#1}}
\newcommand{\mvecerate}[2]{\vec{P}_{\text{xor}={#2}|{#1}}}

\newcommand{\shapeinputswpi}[1]{\dvec_{\labelswapping,{#1}}}
\newcommand{\shapeaftersres}[2]{\dvec_{\text{h}_{#1}|\text{bsm}={#2}}}
\newcommand{\probsres}[1]{p_{\text{bsm}={#1}}}

\newcommand{\tmatx}{\transmat^{\paulix}}

\newcommand{\tmatz}{\transmat^{\pauliz}}

\newcommand{\tmatbitflip}[1]{\transmat^{\text{bf}({#1})}}
\newcommand{\tmatphaseflip}[1]{\transmat^{\text{pf}({#1})}}
\newcommand{\tmatdepol}[1]{\transmat^{\genmatrix{dp}({#1})}}

\newcommand{\tmatdecoh}[1]{\transmat^{\text{sto}({#1})}}

\newcommand{\tmatmem}{\transmat^{\labelmemory}}
\newcommand{\tmatswp}{\transmat^{\labelswapping}}
\newcommand{\mvecswp}{\vec{P}_{\labelswapping}}

\newcommand{\stabiliser}{\mathcal{S}}
\newcommand{\destabiliser}{\mathcal{D}}
\newcommand{\hilbert}[1]{\mathcal{H}_{{#1}}}
\newcommand{\pgroup}[1]{\genset{P}_{{#1}}}

\begin{document}

\title{
On the design of sequential quantum repeaters with error detection
}
\author{Javier \surname{Rey-Domínguez}}
\email{eljrd@leeds.ac.uk}
\affiliation{
School of Electronic and Electrical Engineering, University of Leeds, Leeds LS2 9JT, United Kingdom
}

\author{Koji \surname{Azuma}}
\affiliation{
Basic Research Laboratories, NTT Inc.,
3-1 Morinosato Wakamiya, Atsugi, Kanagawa 243-0198, Japan
}
\affiliation{
NTT Research Center for Theoretical Quantum Physics,
NTT Inc., 3-1 Morinosato-Wakamiya, Atsugi 243-0198, Japan
}

\author{Mohsen \surname{Razavi}}
\affiliation{
School of Electronic and Electrical Engineering, University of Leeds, Leeds LS2 9JT, United Kingdom
}


\begin{abstract}

Long-distance, trust-free quantum communications rely on the development of scalable quantum repeater schemes.
Sequential entanglement generation with error detection is a promising approach that combines manageable implementation requirements with a hop-by-hop architecture compatible with existing classical infrastructure.
In this paper, we investigate key design choices for such schemes over a repeater chain using quantum key distribution as a benchmark.
We find that simple quantum error-detecting codes, such as repetition codes or quantum parity codes, perform well under near-term hardware assumptions and outperform strategies without any encoding.
In particular, we show that two-qubit repetition codes may well be all that we need to implement in many practical scenarios.
We also examine different architectures for the generation of initial entanglement and assess their merits.
Additionally, we show that moderate improvements in some hardware components can partially mitigate the slow technological progress of others.

\end{abstract}


\maketitle


\section{Introduction}
\label{sec:intro}

Quantum technologies continue to attract widespread attention, thanks to their promising applications~\cite{Gibney_QuantumGoldRush:2019,Brooks_BeyondQuantumSupremacy:2019}.
A key area of focus is the development of a global network of quantum devices, i.e. the so-called quantum Internet~\cite{Kimble_TheQuantumInternet:2008, Wehner_QInternet:2018, Illiano_QInternet:2022, Cao_EvolutionOfQKDtoQInternet:2022, Azuma_FromNetworksToQuInternet:2023}.
However, long-distance quantum communications still faces many challenges.
While networks relying on trusted nodes have already been deployed for scientific and military purposes~\cite{Dynes_Cambridge:2019, Koduru_Trusted:2020, Chen_IntegratedQNetwork:2021, Alshowkan_Reconfigurable:2021}, a trust-free architecture is preferable for commercial deployments.

The key solution for trust-free quantum communications in terrestrial settings relies on quantum repeaters~\cite{Briegel_ImperfectOps:1998, Munro_Review:2015, Muralidharan_Optimal:2016}.
Such platforms generally work by distributing entanglement as a resource across a quantum network.
This entanglement can then be used by various quantum applications.
Several strategies are commonly employed by repeater schemes to ensure scalability.
These include generating entanglement between adjacent nodes using a \emph{repeat-until-success} (RUS) approach, performing entanglement distillation to probabilistically increase the quality of the entanglement, and/or encoding the distributed entanglement using quantum error correction (QEC) techniques~\cite{Bennett_MixedState:1996,Razavi_FiberBasedRepeaters:2023}.
A specific type of quantum repeater, the so-called encoded repeaters~\cite{Jiang_EncodedQR:2009}, combines the RUS and QEC mechanisms, limiting the use of the latter to the swapping stage.
This approach can potentially achieve higher rates at long distances when compared with schemes relying exclusively on RUS and distillation~\cite{Briegel_ImperfectOps:1998,Duan_Long:2001}, while remaining less technologically demanding than the so-called one-way schemes that employ QEC to encode and transmit quantum information directly~\cite{Munro_NoMemories:2012,Muralidharan_Ultrafast:2014}.
Nevertheless, longer distances often require encoded repeaters to make use of increasingly complex codes in order to correct errors, which may impede the development of large commercial networks in the short term.

An alternative approach that aims to simplify encoded repeaters to increase their feasibility is to use techniques from QEC exclusively for detecting errors~\cite{Jing_QKDoverQR:2020,Jing_Simple:2021,Shi_MeasurementBased:2025,ReyDominguez_Rethinking:2026}.
When using \emph{error detection}, as opposed to error correction, a system checks the error syndromes only to verify that they correspond to the identity error, and either aborts or restarts operation when this is not the case.
This verification step is generally an easier task than error correction, and benefits from the fact that smaller codes can detect more errors than they can correct.
In the context of entanglement distribution, the \emph{sequential entanglement generation with error detection} (SEG-ED) protocol introduced in \cite{ReyDominguez_Rethinking:2026} executes the error detection step at every intermediate node, involving the next node in the path from one user to the other only if no error was detected within that distribution round.
Subsequently, whenever an error is detected any node beyond the one detecting the error never commits resources to that round, possibly leading to a more efficient use of network resources~\cite{ReyDominguez_Rethinking:2026}.
However, there are still many open questions regarding the performance of this protocol.

In this paper, we remain within the two-user scenario and delve into some of the outstanding design considerations for practical SEG-ED implementations.
As in \cite{ReyDominguez_Rethinking:2026}, we analyse the protocol from the perspective of a repeater chain setup representing a single path between two users within a broader network, and adopt quantum key distribution (QKD)~\cite{Scarani_Security:2009,Lo_SecureQKD:2014,Xu_SecureQKD:2020,Pirandola_Review:2020, Razavi_IntroToQuCommunicationsNetworks:2018} as the benchmark application.
However, in contrast to previous work which attempted to validate SEG-ED's performance by weighing it against other alternatives, here we focus solely on SEG-ED systems and compare their performance under different designs and constraints.
Specifically, we investigate whether larger codes can improve performance under realistic hardware assumptions, explore which method for the generation of initial entanglement fits best depending on the underlying infrastructure, and examine the hardware requirements which enable practical key distribution at long distances.

To address these questions, we expand and generalise the analytical framework from \cite{ReyDominguez_Rethinking:2026}, originally proposed as a tailor-made solution for systems using a three-qubit repetition code.
Using tools from the stabiliser formalism~\cite{Gottesman_HeisenbergRepresentation:1998, Aaronson_ImprovedSimulationOfStabilizer:2004}, we develop a general framework which offers fast and exact calculations of density matrices whenever only stabiliser circuits and Pauli-diagonal noise channels are involved.
Though designed independently, this framework is analogous to a specialisation of Yoder's formalism in \cite{Yoder_GeneralizationOfTheStabilizer:2012}, and can be easily adapted for use in diverse applications.
Here, we show how to use this framework to model encoded repeater chains using CSS codes~\cite{Calderbank_CSS:1996,Steane_CSS:1996}, as these are the most commonly used codes for those systems~\cite{Jiang_EncodedQR:2009}.

The main findings of this paper are summarised below:
\begin{itemize}    
    \item We derive the secret key rates achievable using different error-detecting codes of small size.
    We show that the two-qubit repetition code achieves the highest rates at short-to-moderately-long distances under our near-term hardware assumptions, despite being unable to correct any errors.
    Additionally, we demonstrate that the \emph{Leung-Nielsen-Chuang-Yamamoto} (LNCY) code~\cite{Leung_FourQubitCode1:1997} and the so-called ``four-qubit code"~\cite{Vaidman_ErrorPreventionC422:1996}, i.e. the shortest codes capable of simultaneously detecting bit-flip and phase-flip errors, can outperform the former at long distances and extend the reach of secure communications.

    \item We compare the practical range of the repeater chain using two different configurations for two-photon interference schemes~\cite{Simon_MeetInTheMiddle:2003,Simon_PhotonPairSources:2007,Moehring_EntanglementSingleAtom:2007,Bernien_TwoPhotonNVcenters:2012} for generating the initial entanglement.
    In particular, we compare a setup performing the optical Bell state measurement~\cite{Calsamiglia_MaximumEfficiency:2001,Simon_MeetInTheMiddle:2003,Jones_Design:2016,Bianchi_BSMinQuantumOptics:2026} (BSM) within the memory-equipped repeater nodes to another that requires an additional intermediate node in between each pair of adjacent memory-equipped nodes for this purpose.
    Our results suggest that the former performs better in networks where the devices are further apart, such as current long-distance optical infrastructure~\cite{Hazell_UnderseaFiberSubPlant:2002, Feuerstein_FiberFieldMeasurements:2005}, while the latter setup is better suited to more tightly spaced networks.

    \item We explore how the poor performance of one of the hardware parameters, e.g. due to slow technological development, could be mitigated by the over-performance of some other parameter.
    Specifically, we find that practical key distribution (\(\keyrate > \SI{1}{bps}\)) could be achieved beyond 1000~km with an initial fidelity as low as 95\%, provided around 30 multiplexed quantum channels are available at each hop and the ``four-qubit code" is used.
    
\end{itemize}

The remainder of the paper is structured as follows.
First, \cref{sec:protocol} introduces the setup considered in this paper, as well as how SEG-ED fits into it.
Then, \cref{sec:analysis} explains the mathematical framework used to model such a setup and includes the derivations of the performance metrics.
This analysis is used in \cref{sec:results} to obtain the main results of this paper, based on some realistic parameter values introduced in the same section.
Finally, \cref{sec:conclusion} summarizes the paper and ends with a brief discussion on future directions for this work.

\section{Setup description}
\label{sec:protocol}

\begin{figure}
    \centering
    \includegraphics[width=0.99\columnwidth]{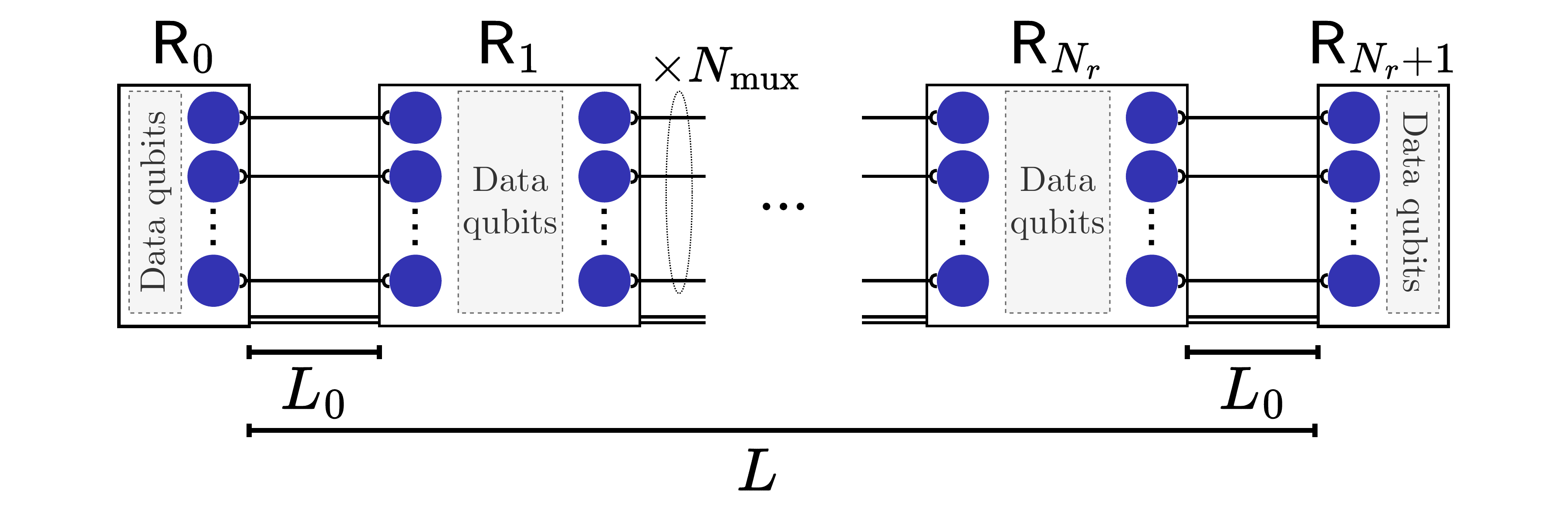}
    \caption{
    Schematic representation of the homogeneous repeater chain.
    Circles represent communication qubits, single lines represent quantum channels and double lines represent classical channels.
    }
    \label{fig:repeater_chain}
\end{figure}

We consider a homogeneous repeater chain between two users, Alice and Bob, separated by a total distance \(\totaldist\).
The chain consists of \(\numrepeaters\) evenly-spaced and identical repeater nodes connected to their two nearest neighbours through physical channels of length \(\hopdist = \totaldist/(\numrepeaters+1)\).
This setup is illustrated in \cref{fig:repeater_chain}, where we have denoted the nodes by \(\repeater{0}\) (Alice), \(\repeater{1}\), \(\repeater{2}\), \dots, and \(\repeater{\numrepeaters+1}\) (Bob).

We assume that each pair of consecutive nodes is connected by a classical channel and \(\numchannels\) multiplexed quantum channels.
Each quantum channel connects a pair of quantum memories, one in each node, allowing them to distribute entanglement through some photon-mediated scheme.
We refer to these quantum memories by the name of \emph{communication qubits}.
In addition to communication qubits, each node also possesses several quantum memories that do not interact directly with the quantum channels, but only with each other and with the communication qubits.
These memories, which we call \emph{data qubits}, fulfil two main roles:
(1)~they participate in the generation of encoded entangled states;
and (2)~they are used for long-term buffering of the encoded data, crucial in sequential schemes~\cite{ReyDominguez_Rethinking:2026}.

We consider that Alice and Bob use the repeater chain to distribute a secret key.
Specifically, they use a \(k\)-qubit variant of BB84-QKD~\cite{Bennett_BB84:1984, Sheridan_SecProofForQKDUsingQudits:2010}.
In order to enable this protocol, which we describe in detail in \cref{sec:multiqubit_qkd}, the repeater chain follows the SEG-ED~\cite{ReyDominguez_Rethinking:2026} strategy using some \([[n,k]]\) CSS code~\cite{Calderbank_CSS:1996,Steane_CSS:1996,Jiang_EncodedQR:2009} to sequentially distribute \(k\)-qubit logical states, encoded using \(n\) physical qubits.
That is, we start from node \(\repeater{0}\) and generate subsequent encoded entangled states, hop-by-hop.
At every hop, we apply a logical BSM to perform the entanglement connection.
If we detect an error during this step, the protocol is aborted; otherwise, we continue until we reach Bob.

In the remainder of this section, we review and discuss this strategy in the context of our setup.
First, we describe the QKD protocol executed by the users in \cref{sec:multiqubit_qkd};
then, we explain the procedure employed at each hop to generate logical entanglement in \cref{sec:logical_gen};
and, finally, we put everything together in \cref{sec:seged}, where we delve into the details of the protocol and how subsequent rounds are initiated.

\subsection{Two-basis multi-qubit QKD}
\label{sec:multiqubit_qkd}

In this paper, we go beyond the standard BB84 protocol~\cite{Bennett_BB84:1984}, and instead consider the protocol described in \cite{Sheridan_SecProofForQKDUsingQudits:2010} to distribute \(k\) key bits per round.
More specifically, we employ the two-basis protocol with the mutually unbiased bases \(\zbasis^{\otimes k}\) and \(\xbasis^{\otimes k}\), and distribute \(k\)-qubit states each round, which are analogous to using \(2^k\)-dimensional qudits instead.
This allows us to maximise the performance of the setup when using error detection codes with \(k > 1\).
The reason is that, if the errors affecting the qubits within a given round are asymmetric (i.e. they affect some qubits differently than others), we obtain a full characterisation of the channel, unlike with the standard BB84 QKD, which strictly focuses on the overall error rate.
We now describe the protocol employed.

Let \(\ket{\physicalbell{k}}_{\system{AB}}\) denote a maximally entangled bipartite state between \(k\)-qubit systems \(\system{A}\) and \(\system{B}\).
This state can be written as:
\begin{equation}\label{eq:physical_bell}
    \Ket{\physicalbell{k}}_{\system{AB}} =
    \frac{1}{2^{k/2}} \sum_{s = 0}^{2^k-1}
    \Ket{s}_{\system{A}} \otimes
    \Ket{s}_{\system{B}} ,
\end{equation}
where \(s\) in \(\ket{s}\) is expressed as a \(k\)-bit string, such that \(\ket{s}\) is a state in the computational basis.

In the entanglement-based version of the protocol~\cite{Sheridan_SecProofForQKDUsingQudits:2010}, Alice prepares such a maximally entangled state each round, and sends one of the \(k\)-qubit systems to Bob, such that they ideally share the state \(\ket{\physicalbell{k}}_{\repeater{0}\repeater{\numrepeaters+1}}\).
Then, they each choose one of \(\zbasis\) or \(\xbasis\) at random, and obtain respective \(k\)-bit strings which we denote by \(\mathsf{k}_{A}\) and \(\mathsf{k}_{B}\).
After many of these rounds have been completed, Alice and Bob discuss the bases they have selected each round over an authenticated classical channel, and discard the rounds on which they chose different bases.
Then, they use a subset of the sifted rounds for parameter estimation, computing the parameters \(\erate{\bbasis}{0}\) through \(\erate{\bbasis}{2^k-1}\), where \(\bbasis \in \{\zbasis, \xbasis\}\) and \(\erate{\bbasis}{\mathsf{x}}\) denotes the probability that \(\mathsf{k}_{A} \oplus \mathsf{k}_{B} = \mathsf{x}\), with \(\oplus\) representing the bitwise XOR operation and \(\mathsf{x}\) being represented as a \(k\)-bit string, provided that both have measured their qubits in the basis \(\bbasis\).
The bit-strings from the remaining rounds go on to be processed by the latter steps of the QKD protocol~\cite{Bennett_BB84:1984,Sheridan_SecProofForQKDUsingQudits:2010}, namely error correction and privacy amplification, before the final key is obtained.

In the context of our setup, we use an encoded version of this protocol.
That is, rather than sharing the state \(\ket{\physicalbell{k}}\), we distribute its logical counterpart, \(\ket{\logicalbell}\), which can be written as:
\begin{equation}\label{eq:logical_bell}
    \Ket{\logicalbell}_{\system{AB}} =
    \frac{1}{2^{k/2}} \sum_{w \in \codespace}
    \Ket{w}_{\system{A}} \otimes
    \Ket{w}_{\system{B}} ,
\end{equation}
for arbitrary \(n\)-qubit systems \(\system{A}\) and \(\system{B}\), where \(\codespace\) denotes a basis for the state space associated to the \([[n,k]]\) code used by the repeater system (note that \( |\codespace| = 2^k\), so the state is normalized).
The users can then apply appropriate decoding measurements to each of the \(n\)-qubit systems in order to obtain the \(k\)-bit strings \(\mathsf{k}_{A}\) and \(\mathsf{k}_{B}\).
In the particular case of CSS codes, these measurements may consist simply of \(n\) single-qubit measurements in the selected basis, plus some conditioning on the set of accepted outcomes.

\subsection{Logical entanglement generation}\label{sec:logical_gen}

Consider two consecutive nodes in the repeater chain, say \(\repeater{i}\) and \(\repeater{i+1}\).
The goal of the logical entanglement generation step is to, ideally, prepare the logical entangled state \(\ket{\logicalbell}_{\repeater{i}\repeater{i+1}}\), defined in \cref{eq:logical_bell}.
To make our analysis more straightforward, we make two important assumptions about the logical entanglement generation procedure:
(1)~It is always initiated on demand; that is, each upstream node, \(\repeater{i}\), must send an explicit initialization message to its downstream neighbour, \(\repeater{i+1}\), in order to request the generation of a logical entangled state.
(2)~It always starts from scratch; that is, if any leftover entanglement resources remain available to the nodes at the end of the procedure, these are discarded before the next request starts.

The logical entanglement generation procedure is composed of two steps: initial entanglement generation and encoding.
After the downstream node receives the initialization message, both nodes employ some probabilistic scheme to entangle \(n\) pairs of communication qubits.
Generally, such a probabilistic scheme allows communication qubits to interact over the quantum channel in order to make an entangling attempt.
The result of the attempt is classically announced to the nodes to verify whether it was successful.
We assume that the nodes use all available quantum channels in parallel, such that these are bundled in ``batches".
In the first batch, all \(\numchannels\) quantum channels attempt to generate entanglement at the same time.
If less than \(n\) pairs of qubits succeed in becoming entangled, those that have failed make another attempt in the following batch.
This goes on until \(n\) (or more) successes are registered.

\begin{figure}
    \centering
    \includegraphics[width=0.99\columnwidth]{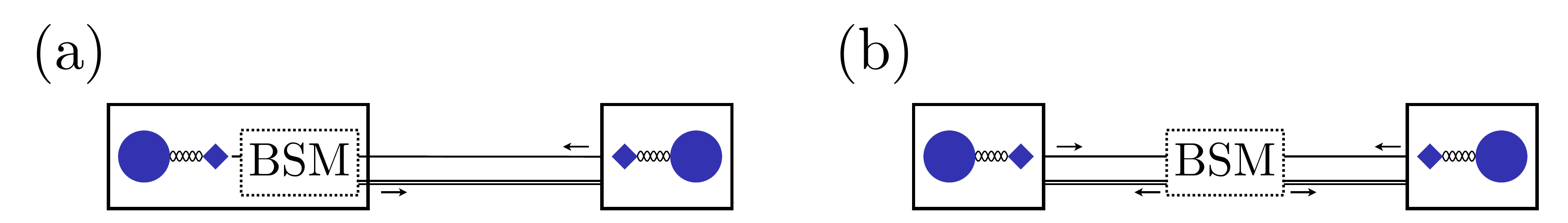}
    \caption{
    Layouts for two-photon interference schemes according to the location of the BSM module.
    (a) Receiver-sender (RxTx).
    (b) Meet-in-the-middle (MITM).
    In both figures, circles represent communication qubits;
    diamonds, photons;
    single lines, quantum channels;
    and double lines, classical channels.
    Arrows indicate the direction of communications.
    }
    \label{fig:bsm_layouts}
\end{figure}

We assume that the following two-photon interference setup~\cite{Simon_MeetInTheMiddle:2003,Simon_PhotonPairSources:2007,Moehring_EntanglementSingleAtom:2007,Bernien_TwoPhotonNVcenters:2012} is used to implement the entanglement generation attempts.
Consider a single pair of communication qubits, connected by a quantum channel.
At each entanglement generation attempt, the communication qubits act as emissive quantum memories~\cite{Duan_Long:2001,Heshami_QuantumMemories:2016,Lei_QuantumOpticalMemory:2023}, becoming entangled to single photons that are then transmitted through the quantum channel, as depicted in \cref{fig:bsm_layouts}.
These photons are interfered in an optical BSM, and the results of the measurement are sent back to the nodes.
If the BSM was successful, the communication qubits are ideally projected into an entangled state.
In these types of setups, the BSM is executed in a device we call the BSM module, also known as a Bell state analyser~\cite{Calsamiglia_MaximumEfficiency:2001,Simon_MeetInTheMiddle:2003,Jones_Design:2016,Bianchi_BSMinQuantumOptics:2026}.
The location of the BSM module determines the timing of the scheme.
We consider here two common options~\cite{Jones_Design:2016}:
(1)~In the receiver-sender (RxTx) layout, depicted in \cref{fig:bsm_layouts}(a), the BSM module is co-located with the upstream node, such that entangled photons are sent upstream and the heralding message travels downstream.
(2)~In the meet-in-the-middle (MITM) layout, depicted in \cref{fig:bsm_layouts}(b), the BSM module is located at the midpoint between the nodes, such that the nodes send photons to it from both directions, and the heralding message then flows outwards.
In practice, each of these layouts has its advantages and disadvantages, which we discuss in \cref{sec:topology_comparison}.

Once the necessary initial entanglement has been distributed, the nodes \(\repeater{i}\) and \(\repeater{i+1}\) respectively prepare \(n\) of their data qubits into the ``logical plus" and ``logical zero" states, denoted by \(\ket{\logicalplus}\) and \(\ket{\logicalzero}\).
Then, using the prepared entangled pairs, they implement a remote logical CNOT gate using local CNOT gates and single-qubit measurements~\cite{Jiang_EncodedQR:2009}.
The results of these measurements, which determine the Pauli frame of the resulting state, are classically exchanged between the nodes.
After appropriate local adjustments, the data qubits in the nodes are projected into \(\ket{\logicalbell}_{\repeater{i}\repeater{i+1}}\).

\subsection{QKD over the SEG-ED repeater chain}
\label{sec:seged}

Let us start the description of our protocol by observing a single key distribution round.
Alice initiates each round by distributing logical entanglement to \(\repeater{1}\), following the method in \cref{sec:logical_gen}.
Then, she immediately selects either the \(\zbasis\) or \(\xbasis\) basis at random, and independently measures all of her \(n\) recently entangled data qubits in said basis.
Using this decoding measurement~\cite{Jing_Simple:2021}, Alice can, on reception of the classical description of the Pauli frame from \(\repeater{1}\), determine whether an error has been detected.
If an error was indeed detected, she lets \(\repeater{1}\) know and aborts the distribution round.
Otherwise, she obtains the \(k\)-bit string \(\mathsf{k}_A\) and the data qubits in \(\repeater{1}\) are projected into a logical state which is completely determined by those bits and the chosen basis.

\begin{figure}
    \centering
    \includegraphics[width=0.99\columnwidth]{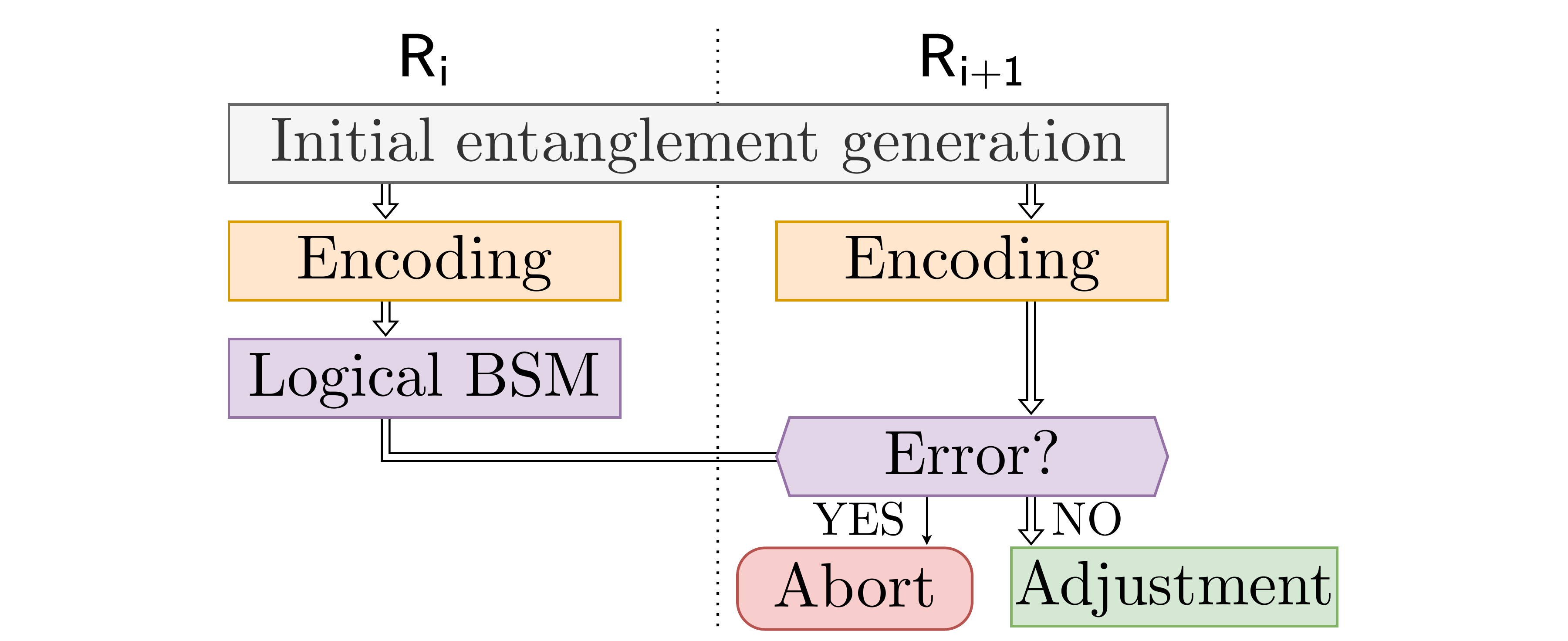}
    \caption{
    Sequence diagram for the operations corresponding to an intermediate hop using the SEG-ED strategy.
    The dotted line is included to denote physical separation.
    Double lines denote a classical transmission of information.
    }
    \label{fig:protocol_description}
\end{figure}

From here, the SEG-ED strategy~\cite{ReyDominguez_QCNC:2025,ReyDominguez_Rethinking:2026} is employed to teleport the encoded data to Bob hop by hop.
We now outline the steps followed at each hop, summarized in \cref{fig:protocol_description}.
Consider that some node, say \(\repeater{i}\), has the encoded data stored in \(n\) of its data qubits.
This node then initiates the logical entanglement generation procedure with the next node in the chain, \(\repeater{i+1}\).
Immediately after applying its local encoding circuit, node \(\repeater{i}\) applies a logical BSM to the newly entangled data qubits and the stored encoded data.
In the case where CSS encoding is employed, this consists of \(n\) transversal CNOT gates, as well as \(2n\) single-qubit measurements~\cite{Jiang_EncodedQR:2009} (see Appendix \ref{app:analysis_appendix}).
The results from the encoding and the logical BSM are then transmitted to \(\repeater{i+1}\), which appends them to its own encoding results and determines whether an error is detected.
If it was, the distribution round is aborted; otherwise, a Pauli frame adjustment recovers the state containing the encoded data in \(\repeater{i+1}\).

Provided no errors are detected, the encoded data is eventually teleported to Bob.
At this point, immediately after his encoding circuit is applied, Bob chooses one of the bases at random and performs his own decoding measurement.
Once he receives the corresponding results from \(\repeater{\numrepeaters}\), which he uses to adjust the Pauli frame of his own observation, Bob checks whether any error was detected by either the last logical BSM or by his own decoding.
If he detects an error, he simply discards the results, effectively aborting the distribution.
Otherwise, he obtains \(\mathsf{k}_B\).
The distribution round is then successfully completed, and the generated bits are employed as described in \cref{sec:multiqubit_qkd}.

To maximize the overall rate, Alice initiates subsequent rounds of the protocol, as described above, as soon as possible.
That is, each round starts as soon as the first hop of the previous round is completed.
We assume every intermediate node has a large enough buffer space (i.e. data qubits) to ensure requests do not accumulate, such that a continuous pipeline of distribution rounds is implemented~\cite{ReyDominguez_Rethinking:2026}.

\section{Rate analysis of the SEG-ED chain}
\label{sec:analysis}

The main performance metric in QKD protocols is the secret key rate, \(\keyrate\).
We compute this in the asymptotic regime and in the absence of eavesdropping as:
\begin{equation}
    \keyrate =
    \secretfrac \distribrate ,
\end{equation}
where \(\secretfrac\) is the secret fraction, representing the expected number of secret key bits that can be extracted from each successful round~\cite{Scarani_Security:2009, Sheridan_SecProofForQKDUsingQudits:2010},
and \(\distribrate\) is the round completion rate, that is, the rate at which a key distribution round is successfully completed.

In the asymptotic regime, the secret fraction can be computed as~\cite{Sheridan_SecProofForQKDUsingQudits:2010}:
\begin{equation}
\label{eq:secret_fraction}
    \secretfrac =
    k
    - \operatorname{H}(\vecerate{\zbasis})
    - \operatorname{H}(\vecerate{\xbasis}) ,
\end{equation}
where:
\begin{equation}
    \operatorname{H}(\vecerate{\bbasis}) :=
    -\sum_{\mathsf{x} = 0}^{2^k-1} \erate{\bbasis}{\mathsf{x}} \log_2(\erate{\bbasis}{\mathsf{x}}) ,
\end{equation}
for \(\bbasis \in \{\zbasis, \xbasis\}\), and the parameters \(\erate{\bbasis}{\mathsf{x}}\) were defined in \cref{sec:multiqubit_qkd}.

From now on, we refer to the parameters \(\erate{\bbasis}{\mathsf{x}}\) above as the error rates of the protocol, though it should be noted that \(\erate{\bbasis}{0}\) corresponds to the case where no bit error is observed.
These rates are derived by introducing a virtual entanglement distribution protocol~\cite{ReyDominguez_Rethinking:2026} between Alice and Bob.
The virtual protocol proceeds in the same manner as described in \cref{sec:protocol}, except that Alice delays her decoding measurement until the end of the round.
Crucially, the virtual protocol is identical to the real protocol provided Alice's data qubits suffer no memory decoherence~\cite{ReyDominguez_Rethinking:2026}.
We analyse this protocol in \cref{sec:main_derivation}.

Next, in order to compute the round completion rate, we assume that a sufficient number of data qubits are allocated to the intermediate nodes to allow for pipelining (see \cref{sec:seged}) without any blockage.
That is, the amount of requests waiting in the buffers does not indefinitely increase. 
Under this assumption, we have that
\begin{equation}
    \distribrate =
    \frac{\probendtoend}{\thop} ,
\end{equation}
where \(\thop\) is the average time required for one hop to process and forward a distribution request, and \(\probendtoend\) is the probability that a distribution round is completed.

We assume that local operations (gates, measurements and classical processing) incur no delay, such that \(\thop := \tgensuccess{n}\), where \(\tgensuccess{j}\) denotes the average time required to entangle \(j\) pairs of communication qubits.
This quantity is computed as:
\begin{equation}
    \tgensuccess{j} =
    \timeinit + \timeattempt \avebatches{j},
\end{equation}
where \(\timeinit\) is the time taken to transmit the initialization message downstream,
\(\timeattempt\) is the duration of an entanglement generation attempt,
and \(\avebatches{j}\) is the average number of batches of attempts (see \cref{sec:logical_gen}) it takes to entangle \(j\) pairs of communication qubits.

The values for \(\timeinit\) and \(\timeattempt\) are determined by the location of the BSM module.
More precisely, \(\timeinit = \hopdist / \lightspeed\) for both the RxTx and MITM layouts, while \(\timeattempt = 2\hopdist / \lightspeed\) for the former and \(\timeattempt = \hopdist/\lightspeed\) for the latter.
The value of \(\avebatches{j}\)~\cite{ReyDominguez_Rethinking:2026}, on the other hand, is:
\begin{equation}\begin{split}
\label{eq:average_batches}
    \avebatches{j} =
    \sum_{t=0}^{j-1} \sum_{l=0}^t \binom{\numchannels}{t} \binom{t}{l} \frac{ (-1)^{t-l} }{ 1 - (1 - \probgen)^{\numchannels-l} } ,
\end{split}\end{equation}
where \(\probgen\) is the probability that a pair of communication qubits becomes entangled after one attempt, calculated as:
\begin{equation}\label{eq:probgen}
    \probgen =
     \frac{(\probcou \detefficiency)^2}{2} 10^{-\hopdist\chatten/10} ,
\end{equation}
with \(\probcou\) denoting the effective coupling efficiency of each quantum memory to the fibre~\cite{ReyDominguez_Rethinking:2026}, including any losses related to frequency conversion;
\(\detefficiency\) denoting the efficiency of the single photon detectors in the BSM;
\(\chatten\) denoting the attenuation parameter of the fibre;
and the factor of 2 accounting for the fact that only two Bell states may be distinguished if the BSM consists exclusively of linear optics and photodetectors~\cite{Calsamiglia_MaximumEfficiency:2001}.

The remainder of the section is structured as follows.
First, we describe the error models we consider for our setup in \cref{sec:error_models}.
Then, we introduce in \cref{sec:framework} some background on stabiliser and CSS codes, and review the framework we use to analyse the virtual protocol.
Finally, we use a recursive approach to complete the analysis in \cref{sec:main_derivation}.

\subsection{Error models}
\label{sec:error_models}

We account for three main sources of noise in the repeater chain:
(1)~errors in the initial entanglement between the communication qubits;
(2)~circuit errors, including measurement and gate errors;
and, (3)~ memory decoherence.
We now provide the formal descriptions for the relevant noise channels.

We assume that, at the moment that two nodes receive confirmation of a pair of their respective communication qubits becoming entangled, the state in those qubits can be expressed as a Werner state with fidelity \(\initfid\) with respect to \(\ket{\phi^+} = (\ket{00}+\ket{11})/\sqrt{2}\).
Specifically, this initial state is described by a density matrix \(\initstate\) of the following form:
\begin{equation}
\label{eq:initial_bell}
    \initstate =
    \initfid \Ketbra*{\phi^+}
    + (1 - \initfid) \frac{\identity_4 - \Ketbra*{\phi^+}}{3} ,
\end{equation}
where \(\identity_4\) denotes the \(4\times4\) identity matrix.

All circuits considered in this paper consist only of single-qubit measurements, as well as single-qubit and two-qubit gates.
We assume that each single-qubit measurement has an independent probability of error \(\probmeaserr\).
In contrast, we consider single-qubit gates to be ideal.
This is because these gates are often less error-prone than two-qubit gates~\cite{Brown_SingleQubitGateError10-4:2011,Harty_HighFidelityPreparation:2014,Harty_HighFidelityTrappedLogic:2016}, leading to an almost negligible contribution to the overall noise.
For two-qubit gates, on the other hand, we assume that the pair of qubits on which the gate acts undergo a joint depolarizing channel with probability \(\probgateerr\).
That is, consider some gate acting on qubits \(\system{A}\) and \(\system{B}\).
Provided the ideal operation of the gate is described by the unitary matrix \(U_{\system{AB}}\), the action of the gate on a quantum state with density matrix \(\rho\) is described by the following transformation:
\begin{equation}\begin{split}
\label{eq:twoqubit_gate}
    \rho \to
    (1 - \probgateerr) U_{\system{AB}} \rho U_{\system{AB}}^\adjoint
    + \frac{\probgateerr}{4} \ptrace{\system{AB}} \{\rho\} \otimes \identity_{\system{AB}} ,
\end{split}\end{equation}
where \(\ptrace{\system{AB}}\) denotes the partial trace on qubits \(\system{A}\) and \(\system{B}\), and \( \identity_{\system{AB}}/4 \) corresponds to the maximally mixed state in those qubits.

Lastly, we consider two models for the memory decoherence: \emph{dephasing} and \emph{depolarizing}.
Suppose a qubit, \(\system{Q}\), is stored for some time \(\tau\) in a quantum memory with coherence time \(\coherencetime\).
Under the dephasing model, \(\system{Q}\) suffers a phase flip with a probability \([1 - \exp(-\tau/\coherencetime)]/2\), and remains untouched otherwise.
That is, given an initial state with density matrix \(\rho\), the decoherence is expressed via the quantum operation \(\qchandephase{\tau}_{\system{Q}}\), defined as:
\begin{equation}
\label{eq:memory_dephasing}
    \qchandephase{\tau}_{\system{Q}}(\rho) =
    \frac{1 + e^{\frac{-\tau}{\coherencetime}} }{2} \rho
    +
    \frac{1 - e^{\frac{-\tau}{\coherencetime}} }{2} Z_{\system{Q}} \rho Z_{\system{Q}} ,
\end{equation}
where \(Z_{\system{Q}}\) denotes the Pauli-Z operator acting on qubit \(\system{Q}\).
Under the depolarizing model, on the other hand, the qubit fully depolarizes (i.e. becomes the maximally mixed state) with probability \(1 - \exp(-\tau/\coherencetime)\).
This transformation is described by the operation \(\qchanmemdepol{\tau}_{\system{Q}}\), defined as:
\begin{equation}\begin{split}
\label{eq:memory_depolarizing}
    \qchanmemdepol{\tau}_{\system{Q}}(\rho) =
    &\frac{1 + 3 e^{\frac{-\tau}{\coherencetime}} }{4} \rho
    +
    \frac{1 - e^{\frac{-\tau}{\coherencetime}} }{4}\Big(
    X_{\system{Q}} \rho X_{\system{Q}}
    \\
    &+ Y_{\system{Q}} \rho Y_{\system{Q}}
    + Z_{\system{Q}} \rho Z_{\system{Q}}
    \Big) ,
\end{split}\end{equation}
where \(X_{\system{Q}}\) and \(Y_{\system{Q}}\) are the Pauli-X and Pauli-Y operators on qubit \(\system{Q}\).  

\subsection{Analytical framework}
\label{sec:framework}

We now review the framework proposed in \cite{ReyDominguez_Rethinking:2026} and outline how we implement and improve it within the context of our analysis.
We provide further details about this implementation in Appendix \ref{app:diagonal_stabiliser}, where we also introduce relevant tools from the stabiliser formalism~\cite{Gottesman_HeisenbergRepresentation:1998, Aaronson_ImprovedSimulationOfStabilizer:2004}.

Consider an arbitrary quantum state with density matrix \(\rho\).
It is possible to find an orthonormal basis \(\bbasis = \{ \ket{b_0}, \ket{b_1}, \dots, \ket{b_{D-1}} \} \)  for its associated state space, where \(D\) is the dimension of the Hilbert space, such that \(\rho\) can be written as:
\begin{equation}\label{eq:diagonal_rho}
    \rho =
    \sum_{i = 0}^{D-1} \lambda_i \Ketbra*{b_i} ,
\end{equation}
for \(\lambda_i \geq 0\) and \(\sum_i \lambda_i = 1\).
That is, any state can be expressed as a \emph{probabilistic mixture} of states in some orthonormal basis, e.g., \(\bbasis\).
We define the \emph{diagonal representation} of the state \(\rho\) to be the tuple \(\{\dvec, \bbasis\}\), where we refer to \(\dvec\) as the \emph{diagonal vector} of the state \(\rho\) in \(\bbasis\), defined as:
\begin{equation}
    \dvec =
    \begin{bmatrix}
        \lambda_0 & \lambda_1 & \dots & \lambda_{D-1}
    \end{bmatrix}^\transpose .
\end{equation}

Whenever a state admits a diagonal representation with respect to a specific basis, we say that the state is diagonal in that basis.
Moreover, we express the diagonal representation of a state by using only its diagonal vector whenever the basis is clear from context.

System composition can be neatly represented in the diagonal representation.
To see this, consider two states, \(\{\dvec_a, \bbasis_a\}\) and \(\{\dvec_b, \bbasis_b\}\), for two orthonormal bases \(\bbasis_a\) and \(\bbasis_b\).
Then, the joint state of the compound system admits a diagonal representation \(\{\dvec_a \kron \dvec_b, \bbasis_a \otimes \bbasis_b\}\), where \(\kron\) denotes the Kronecker product.

Furthermore, the diagonal representation is convenient to represent certain quantum operations.
In particular, consider an operation \(\qoperation\) that results in a diagonal state in some basis, \(\bbasis_{\text{out}} = \{\ket{v_0}, \dots, \ket{v_{D'-1}}\}\), whenever the input is diagonal in some basis, \(\bbasis_{\text{in}} = \{\ket{b_0}, \dots, \ket{b_{D-1}}\}\), which may coincide with \(\bbasis_{\text{out}}\).
That is, \(\qoperation(\{\dvec_{\text{in}}, \bbasis_{\text{in}}\}) = \{\dvec_{\text{out}}, \bbasis_{\text{out}}\}\) for any diagonal vector \(\dvec_{\text{in}}\).
Then, the diagonal vector of the output state in \(\bbasis_{\text{out}}\) can be computed as:
\begin{equation}
\label{eq:diagonal_matmul}
    \dvec_{\text{out}} =
    \transmat \dvec_{\text{in}} ,
\end{equation}
where \(\transmat\) is a \emph{transition matrix} of dimension \(D' \times D\), whose elements are given by:
\begin{equation}
\label{eq:transitionmat_elements}
    \transmat_{ij} :=
    \bra{v_i}
    \qoperation(\ketbra*{b_j})
    \ket{v_i} ,
\end{equation}
with \(i, j\) being the row and column indices, respectively.
Whenever \(\qoperation\) is trace-preserving, the additional condition \(\sum_{i} \transmat_{ij} = 1\) must hold for all \(j\), while \(\sum_{i} \transmat_{ij} \leq 1\) holds for all \(j\) in any trace-non-increasing operation.

Importantly, unlike the diagonal representation of a state, the transition matrix \(\transmat\) does not provide a complete description of the operation \(\qoperation\), since it is conditioned on the input state being diagonal in \(\bbasis_{\text{in}}\).
Nevertheless, this partial description is sufficient for many important operations.

Consider now some positive operator-valued measure (POVM) \(\genset{M} = \{\povm_0, \povm_1, \dots, \povm_{M-1}\}\), where outcome \(r\) is associated to the positive operator \(\povm_r\), acting on a diagonal state \(\{\dvec, \bbasis\}\) in orthonormal basis \(\bbasis = \{\ket{b_0}, \dots, \ket{b_{D-1}}\}\).
The probability that outcome \(r\) occurs is given by:
\begin{equation}
\label{eq:diagonal_povm}
    p_r =
    \vec{P}_r^\transpose \dvec ,
\end{equation}
where \(\vec{P}_r\) is the \emph{measurement vector} associated to outcome \(r\) in basis \(\bbasis\), given by:
\begin{equation}
\label{eq:diagonal_povm_elements}
    \vec{P}_r =
    \begin{bmatrix}
        \braket{b_0|\povm_r|b_0} , &
        \dots , &
        \braket{b_{D-1}|\povm_r|b_{D-1}}
    \end{bmatrix}^\transpose .
\end{equation}

The diagonal representation introduced here is particularly convenient when dealing with any combination of \emph{stabiliser circuits}~\cite{Gottesman_HeisenbergRepresentation:1998,Aaronson_ImprovedSimulationOfStabilizer:2004} and \emph{Pauli-diagonal channels}.
Consequently, it is useful for our setting, since CSS codes~\cite{Calderbank_CSS:1996,Steane_CSS:1996} are a subclass of stabiliser codes~\cite{Cleve_EfficientEncodingsForQEC:1997}, and thus the ideal operations of the repeater chain correspond to stabiliser circuits.
In addition, all noise processes in \cref{sec:error_models} are Pauli-diagonal, i.e., for an operation \(\qoperation\) acting on \(m\) qubits in state \(\rho\), the output is given by:
\begin{equation}
\label{eq:pauli_diagonal}
    \qoperation(\rho) =
    \sum_{\sigma \in \pgroup{m}}
    \lambda_\sigma \sigma \rho \sigma^\adjoint ,
\end{equation}
where \(\rho\) is some density matrix;
\(\pgroup{m}\) is the Pauli group on \(m\) qubits;
and \(\sum_{\sigma \in \pgroup{m}} \lambda_\sigma = 1\), with \(\lambda_\sigma \geq 0\).

Based on the above, we note that, for any stabiliser state~\cite{Gottesman_HeisenbergRepresentation:1998,Aaronson_ImprovedSimulationOfStabilizer:2004}, one can construct an orthonormal basis containing that state such that any Pauli operator maps one element in the basis to another, up to a global phase.
This is known as the stabiliser basis~\cite{Yoder_GeneralizationOfTheStabilizer:2012}.
Consequently, the action of any Pauli-diagonal channel can be represented by a transition matrix from this basis to itself.
Moreover, ideal stabiliser circuits can be efficiently simulated by updating the underlying basis~\cite{Yoder_GeneralizationOfTheStabilizer:2012}, while leaving the diagonal vector unchanged.

\subsection{Analysis of the virtual protocol}
\label{sec:main_derivation}

We now derive the error rates in \cref{eq:secret_fraction} and the round completion probability \(\probendtoend\) within the virtual protocol.
First, suppose that Alice and Bob have successfully distributed logical entanglement between them without detecting any errors at any intermediate swap.
Then, before the decoding measurement, their end-to-end entangled state has diagonal representation \(\{\shapeendtoend, \basisent\}\), where \(\basisent\) is the stabiliser basis for the logical entangled state, \(\ket{\logicalbell}\), as defined in \cref{eq:logical_bell}.
Consequently, we compute the error rates as:
\begin{equation}
\label{eq:error_rates}
    \erate{\bbasis}{j}
    =
    \frac{\mvecerate{\bbasis}{j}^\transpose \shapeendtoend}{\probdecode{\bbasis}} ,
\end{equation}
where \(\bbasis \in \{\zbasis, \xbasis\}\);
\(\mvecerate{\bbasis}{j}\) is the measurement vector associated to the outcome where the XOR between Alice's and Bob's key bits is \(j\) provided the round was completed and they chose to measure in \(\bbasis\);
and \(\probdecode{\bbasis}\) is the probability that no error is detected during the decoding under the same conditions.
We compute \(\probdecode{\bbasis}\) as:
\begin{equation}
\label{eq:prob_decode}
    \probdecode{\bbasis} =
    \sum_{j=0}^{2^k-1} \mvecerate{\bbasis}{j}^\transpose \shapeendtoend ,
\end{equation}
and we derive \(\mvecerate{\bbasis}{j}\) in Appendix \ref{app:decoding}.

The value of \(\shapeendtoend\) can be obtained recursively~\cite{ReyDominguez_QCNC:2025,ReyDominguez_Rethinking:2026}.
Specifically, provided our assumption of smooth pipelining holds (see \cref{sec:seged}), we can analyse each hop independently.
Thus, we define \(\{\shapeafteriswaps{i}, \basisent\}\) as the diagonal representation of the state distributed between Alice and \(\repeater{i+1}\) after \(i\) hops, provided no errors were detected.
It follows that \(\shapeendtoend = \shapeafteriswaps{\numrepeaters}\).
Additionally, we set the first logical entangled state, shared between Alice and \(\repeater{1}\), to \(\shapeafteriswaps{0} := \shapeafterencoding\), where \(\{\shapeafterencoding, \basisent\}\) is the state generated via the logical entanglement generation procedure.

In order to simplify our analysis, we make the assumption that the time it takes to entangle \(j\) pairs of communication qubits from the start of the logical communication is given by the expected value \(\tgensuccess{j}\) at every hop~\cite{ReyDominguez_Rethinking:2026}.
Notably, though on average the total entanglement generation time remains unchanged under this simplification, the total decoherence suffered by the qubits does not.
This is because, as shown in \cref{eq:memory_dephasing,eq:memory_depolarizing}, the probability of a qubit maintaining coherence is not linear with respect to the storage time.
Specifically, this probability is a convex function of the storage time, so our simplification overestimates the decoherence suffered by the qubits during the entanglement generation phase.
Additionally, a corollary of the above simplification is that our analysis disregards queuing times, since the timing of every hop in the repeater chain is deterministic.
Naturally, this is not true in a practical setting, as the time spent buffering in packet-switched networks can often be significant.
Nevertheless, an analysis of the queueing times would require a more complex model for the queue system (see e.g. the analysis of an unencoded repeater chain in \cite{Guedes_Sequential:2024}), and important assumptions regarding the traffic of the overall network.
Thus, our simplified model can be interpreted as giving us a view of the repeater chain during a period of relatively low-traffic in the network.

Taking the simplifications above into account, we now compute:
\begin{equation}
\label{eq:shape_recursive}
    \shapeafteriswaps{i} =
    \frac{ \tmatswp \left( \tmatmem \shapeafteriswaps{i-1} \kron \shapeafterencoding \right) }
    { \probhop{i} } ,
\end{equation}
where \(\tmatswp\) is the transition matrix associated to the operation of the logical BSM selecting the outcomes associated to no errors being detected;
\(\tmatmem\) is the transition matrix associated to the memory decoherence of the data qubits in \(\repeater{i}\) for a time \(\thop\);
and \(\probhop{i}\) is the probability that no errors are detected by this operation, calculated as:
\begin{equation}
\label{eq:prob_swap}
    \probhop{i} =
    \mvecswp^\transpose \left( \tmatmem \shapeafteriswaps{i-1} \kron \shapeafterencoding \right) ,
\end{equation}
with \(\mvecswp\) denoting the measurement vector associated to detecting no errors in the logical BSM.
We derive the values of \(\tmatswp\), \(\tmatmem\) and \(\mvecswp\) in Appendix \ref{app:swapping}.

\newcommand{\tmatnoisyenc}{\transmat_{\labelencoding\text{,err}}}
\newcommand{\tmatencbell}[1]{\transmat_{\labelencoding,g_{#1}}}

We now turn our attention to the derivation of \(\shapeafterencoding\), that is, the \(\basisent\)-diagonal vector of the logical entangled state generated at each hop.
As we show in Appendix \ref{app:encoding}, this can be computed using only \(2n\)-qubit operations, such that:
\begin{equation}
\label{eq:shape_afterencoding}
    \shapeafterencoding =
    \transmat_{\labelencoding}
    \left( \shapelogicalplus \kron \shapelogicalzero \right) ,
\end{equation}
where \(\transmat_{\labelencoding}\) is the overall transition matrix representing the encoding operation, and \(\shapelogicalplus\) and \(\shapelogicalzero\) denote the diagonal vectors for the noisy preparations of the data qubits in the upstream and downstream nodes, respectively.
The values of these diagonal vectors are derived in Appendix \ref{app:encoding}, while the transition matrix can be further decomposed into:
\begin{equation}
\label{eq:tmat_encoding}
    \transmat_{\labelencoding} :=
    \tmatnoisyenc
    \prod_{j=1}^n \tmatencbell{j} ,
\end{equation}
where \(\tmatnoisyenc\) is the transition matrix representing the errors in the CNOT gates and the measurements, while \(\tmatencbell{j}\) denotes the transition matrix introducing the errors brought over by the noise in the \(j\)-th pair of communication qubits that became entangled.
Notably, these matrices take into account not only the initial infidelity of the Bell pairs, \(1 - \initfid\), but also the memory decoherence suffered by each pair of communication qubits between the time they became entangled to each other and the time the encoding operation was applied, that is, \(\tgensuccess{n} - \tgensuccess{j}\) under the assumptions introduced before.

Finally, we derive the round completion probability, \(\probendtoend\), as:
\begin{equation}
\label{eq:prob_endtoend}
    \probendtoend =
    \probdecode{\zbasis}
    \prod_{i=1}^{\numrepeaters} \probhop{i} ,
\end{equation}
where we have assumed that the users selected the \(\zbasis\) in most rounds in order to maximise the fraction of rounds contributing to the final key.

\section{Results}
\label{sec:results}

In this section, we discuss three core aspects of the design of SEG-ED repeater chains.
First, in \cref{sec:code_comparison}, we compare the performance of the system using different error detection codes.
Next, in \cref{sec:topology_comparison}, we examine the impact of the BSM module's location and the separation between nodes.
Lastly, we investigate in \cref{sec:fixing_slow} how slow technological progress in one parameter can be compensated by improving other aspects of the system.

\subsection{Comparison between error detection codes}
\label{sec:code_comparison}

Suppose the hardware of the repeater chain is fixed to the parameters in \cref{tab:parameters}, and the decoherence model of the quantum memories is known.
Additionally, the nodes are placed \(\hopdist = \SI{50}{\kilo\meter}\) apart from each other and the BSM modules are used inside each of them, as in the RxTx layout in \cref{fig:bsm_layouts}(a).
These are likely design constraints in early quantum networks, where the breadth of commercially available hardware is limited and deployments must be integrated with pre-existing infrastructure.
In fact, the parameters in \cref{tab:parameters} are meant to offer a realistic view of the capabilities that may be available in the near-to-medium term, with some of them reflecting numbers that have already been achieved in some platforms~\cite{Marsili_DetectingPhotons93:2013,ITUT_SingleModeFiber:2024,Harty_HighFidelityPreparation:2014,Harty_HighFidelityTrappedLogic:2016,Clark_HighfidelityBell:2021,Nemirovsky_TwoQubitGateNeutralAtoms:2021}.
Our goal in this section is to select the error detection code that achieves the best performance within these constraints.

\begin{table}
    \centering

    \begin{equation*}\begin{array}{ccccccccc}
        \toprule
        \probcou & \detefficiency & \chatten \left(\frac{\si{\decibel}}{\si{\kilo\meter}}\right) & \initfid & \probgateerr & \probmeaserr & \coherencetime (\si{\second}) & \numchannels & \lightspeed \left(\frac{\si{\kilo\meter}}{\si{\second}}\right)
        \\ \midrule
        0.4 & 0.9 & 0.15 & 0.99 & 0.001 & 0.001 & 1 & 12 & 200000
        \\ \bottomrule
    \end{array}\end{equation*}
    
    \caption{
    Configuration parameters.
    }
    \label{tab:parameters}
\end{table}

Consider first the case where memory decoherence introduces phase flips, according to the dephasing model in \cref{eq:memory_dephasing}.
Provided memory effects contribute to a significant portion of the overall noise of the system, as is often the case with sequential schemes, errors in the \(\xbasis\) basis will dominate.
Therefore, we focus on error detection codes that mitigate errors in that basis; namely, we focus on the family of \emph{rotated qubit repetition codes}.

The rotated \(n\)-qubit repetition code consists of the codewords \(\ket{+}^{\otimes n}\) and \(\ket{-}^{\otimes n}\).
That is, it is an \([[n,1]]\) code, and it can be shown to be capable of detecting up to \(n-1\) phase-flips.
This family of codes is valued for its simplicity as well as practicality.
Note that the codewords can be easily prepared through \(n\) single qubit rotations, making the logical zero state \(\ket{\logicalzero}\) reasonable to reproduce with high fidelity in many platforms.
Additionally, the logical plus state, \(\ket{\logicalplus}\), corresponds to the rotation of an \(n\)-qubit GHZ state~\cite{Greenberger_GHZ:1989}, whose preparation methods are well-studied across different platforms~\cite{Jin_PreparingGHZOnNeutralAtoms:2025,Anikin_LemniscatePhaseTrajectoriesGHZ:2026,Minnella_SingleGateMultipartiteEntanglementOnRoomTemp:2026}.

From the family of repetition codes, the case \(n=3\) has often been used to benchmark systems based on encoded repeaters~\cite{Jiang_EncodedQR:2009,Jing_QKDoverQR:2020,ReyDominguez_Rethinking:2026}, as it corresponds to the smallest code capable of correcting any phase-flip on a single qubit.
Nevertheless, here we are focusing on error detection, and thus the scheme with \(n = 2\) is also potentially useful.
In fact, the two-qubit repetition code has the advantage that the logical plus state corresponds to a Bell state, making its preparation practical even when considering short-term hardware.
On the other hand, it is also of interest to explore whether using larger repetition codes, with \(n > 3\), can provide any benefit to the performance.

\begin{figure*}
    \centering
    \includegraphics[width=0.99\linewidth]{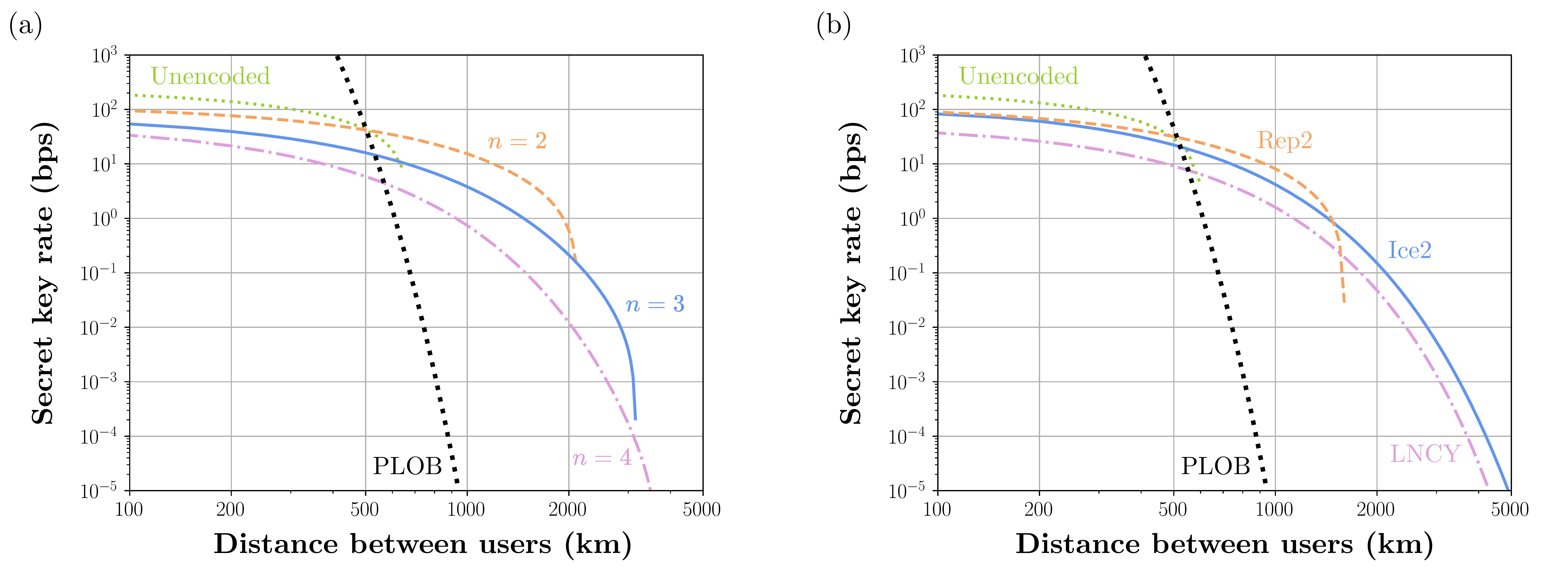}
    \caption{
    Secret key rate achieved against the distance between users, provided the repeater nodes are separated by \(\SI{50}{\kilo\meter}\).
    (a) Rate achieved using repetition codes when memory decoherence follows the dephasing model.
    (b) Rate achieved using different codes when memory decoherence follows the depolarizing model.
    In both figures, we also plot the performance of an unencoded repeater chain~\cite{ReyDominguez_Rethinking:2026} under the parameters summarized in \cref{tab:parameters}, as well as the bound for a repeaterless channel, i.e. the PLOB bound~\cite{Pirandola_PLOB:2017}, at a clock rate of 1~GHz.
    }
    \label{fig:comparing_codes}
\end{figure*}

To find this out, we plot the secret key rate achieved using three repetition codes of different sizes as a function of the distance between users in \cref{fig:comparing_codes}(a).
For reference, we plot the rate of an unencoded solution alongside these, as well as the maximum rate of a repeaterless channel~\cite{Pirandola_PLOB:2017}, i.e. the PLOB bound, which we calculate here as:
\begin{equation}
\label{eq:plob}
    \text{PLOB} := -10^9 \log_2(1 - 10^{-\totaldist\chatten/10}) ,
\end{equation}
where we have considered a hypothetical clock rate of 1~GHz.

\Cref{fig:comparing_codes}(a) reveals that, while the unencoded setup cannot reach distances far beyond 600~km, every encoded system enables positive key rates well above 1000~km.
In particular, while all three of those setups manage to convincingly surpass the repeaterless bound, the two-qubit code (labelled `\(n=2\)' in the plot) achieves the best performance for most of the effective range of the systems.
Specifically, this advantage can be observed up to around 2000~km, where the solution with \(n=3\) briefly becomes comparable before the rates of both solutions vanish.
This is of significant practical importance, since the case \(n=2\) is the most feasible to implement with today's technology, as we previously discussed.
Indeed, this result shows that SEG-ED can enable long-distance, trust-free QKD by relying on very simple and practical codes.

On the other hand, while larger codes do slightly extend the theoretical range at which the secret key rate is positive, the actual values of \(\keyrate\) dip below 0.1~bps well before the two-qubit code ceases to offer the strongest performance, at which point they likely become impractical.
These results suggest that more precise equipment might be needed to take full advantage of larger codes.

Next, we consider the scenario where memory decoherence follows the depolarizing model in \cref{eq:memory_depolarizing}.
In contrast to the previous case, errors now have comparable effects in both bases.
Therefore, while achieving good protection in only one basis can still provide us with a decent performance, thanks to QKD's robustness to errors in a single basis, it is possible that slightly more complex codes offering symmetric performance may be preferable.

We now introduce two new codes fitting such description.
These are the \([[4,1]]\) Leung-Nielsen-Chuang-Yamamoto (\emph{LNCY}) code~\cite{Leung_FourQubitCode1:1997}, and the \([[4,2]]\) so-called ``four-qubit code"~\cite{Vaidman_ErrorPreventionC422:1996,Grassl_CodesForErasureChannel:1997}, which we will refer to as the \emph{Iceberg-2} code~\cite{Self_IcebergCode:2024} for the sake of clarity.
Both of these codes are capable of detecting arbitrary errors (bit-flips or phase-flips) on a single qubit, though they are not capable of correcting any errors.

The LNCY code consists of two codewords, \(\ket{\phi_{0000}}\) and \(\ket{\phi_{0011}}\), where we define
\begin{equation}
\label{eq:bipartite_not_state}
    \ket{\phi_{\mathsf{s}}} :=
    \frac{1}{\sqrt{2}} (\ket{\mathsf{s}} + \ket{\overline{\mathsf{s}}}) ,
\end{equation}
with \(s\) being a binary bit string.
The Iceberg-2 code, on the other hand, consists of both of the LNCY codewords, in addition to \(\ket{\phi_{0101}}\) and \(\ket{\phi_{0110}}\).
Upon inspection of the codewords above, we notice that all of them may be prepared using a four-qubit GHZ state preparation method, plus potentially some single-qubit adjustments.
However, the LNCY code proved to be more practical when looking at the logical plus state, which may be prepared as a product state of two Bell pairs.
This contrasts with the Iceberg-2 code, for which the straightforward preparation is to rotate every qubit in a four-qubit GHZ state.

We plot the secret key rates of SEG-ED chains using these two codes, as well as the two-qubit repetition code, in \cref{fig:comparing_codes}(b).
The first thing we notice is that the two-qubit repetition code (labelled ``Rep2" in the figure) still achieves rates over 1~bps up to distances of around 1400~km, despite its minimal error detection capabilities.
In fact, similar to the previous scenario, this code achieves the highest rates for distances roughly between 500~km and 1400~km.
Unlike before, however, both larger codes still achieve near-practical rates at longer distances.
Specifically, the Iceberg-2 code (labelled ``Ice2" in the figure) achieves \(\keyrate > \SI{0.1}{bps}\) up to 2100~km, while the LNCY code does the same up to 1750~km.
Furthermore, the rate of the setup using the Iceberg-2 code remains close to that of the repetition code for most of the plotted range, positioning this code as a strong contender for scenarios where memory errors are unbiased.
In fact, as expected, it achieves approximately double the secret key rate of the LNCY code at all distances, since twice the logical qubits distributed in each round, although this comes at the cost of a slightly noisier initial state preparation.

\subsection{Impact of the BSM location}
\label{sec:topology_comparison}

\begin{figure}
    \centering
    \includegraphics[width=0.99\linewidth]{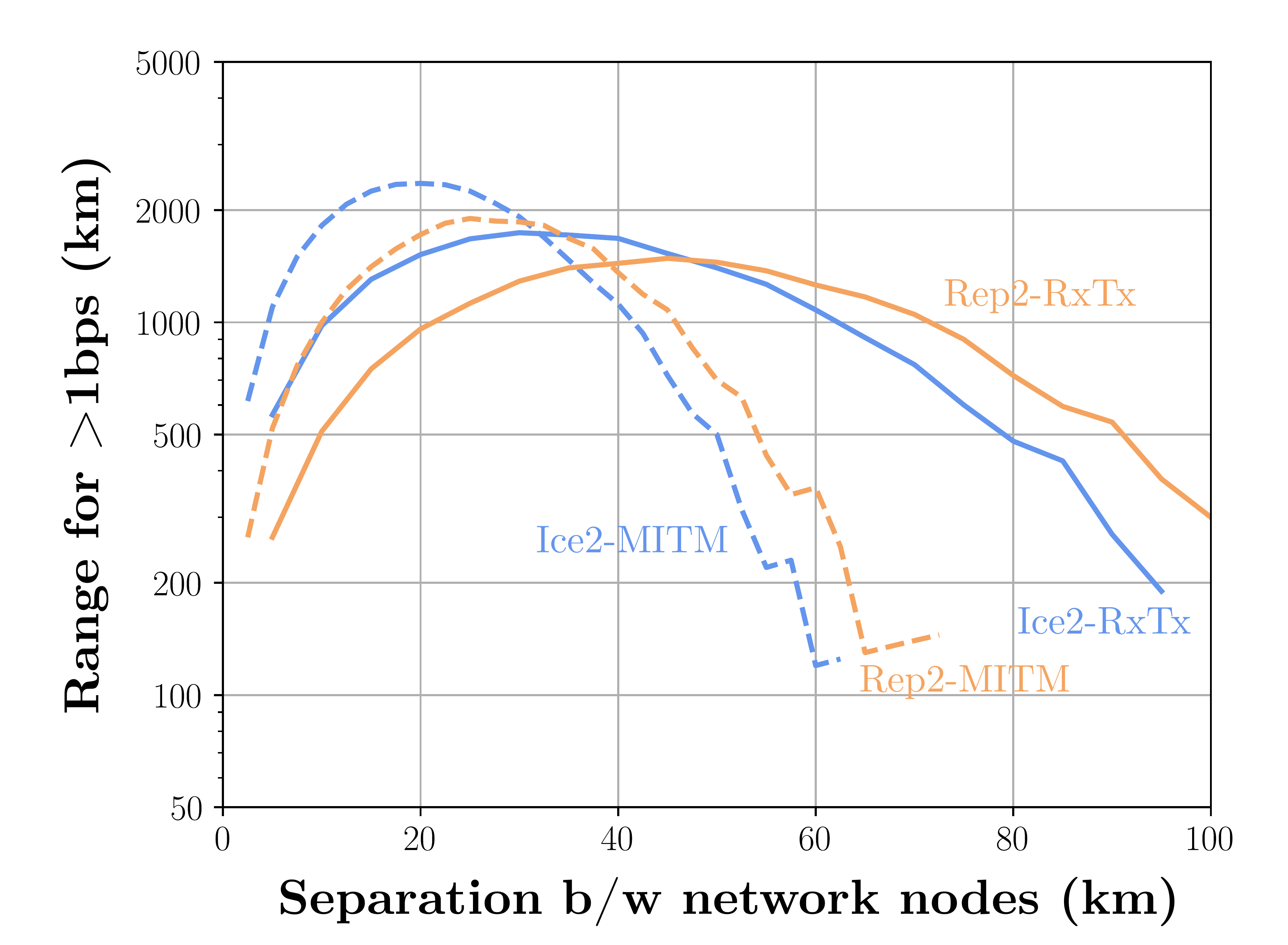}
    \caption{
    Maximum distance at which a repeater chain can maintain \(\keyrate > \SI{1}{bps}\) against the separation between network nodes; that is, \(\hopdist\) for the RxTx layout and \(\hopdist/2\) for the MITM layout.
    Memory decoherence is depolarizing.
    The solid (dashed) lines represent repeater chains using the RxTx (MITM) BSM layout.
    The orange lines correspond to a repeater chain using the two-qubit repetition code (``Rep2"), while the blue lines correspond to a repeater chain using the Iceberg-2 code (``Ice2").
    }
    \label{fig:range_for_layouts}
\end{figure}

Consider now the following scenario.
We want to deploy a quantum network that will be integrated with existing optical fibre infrastructure, but we are not allowed to deploy new nodes.
We may choose either the RxTx or MITM layouts but, in the latter case, the BSM module must be deployed at one of the existing nodes.
That is, the effective separation between repeater nodes is double in the case of the MITM layout.

Such a scenario is interesting when we consider the potentially high initial cost of deploying new nodes.
In fact, the hardware for the BSM is in principle simpler than for a full repeater node, since it does not need quantum memories.
Thus, a system based on the MITM layout can be less costly if it allows the separation between repeater nodes (equipped with memories) to become longer, spreading the repeater nodes further apart while maintaining performance.

To investigate this, we plot in \cref{fig:range_for_layouts} the \emph{practical range}, i.e. the maximum distance at which \(\keyrate > \SI{1}{bps}\), as a function of the separation between network nodes, provided either the RxTx or the MITM layouts are used, and using either the two-qubit repetition code or the Iceberg-2 code.
We assume that the hardware is the same in all cases, and that the quantum memories suffer depolarizing decoherence.

Note that when refer here to network nodes, we refer to any device installed in the network connected to others through fibre.
This is not to be confused with the repeater nodes, which are necessarily equipped with quantum memories and are the ones that apply the encoding and swapping operations.
That is, the separation between network nodes in the RxTx and MITM layouts corresponds, respectively, to \(\hopdist\) and to \(\hopdist/2\).

The first thing that we notice in \cref{fig:range_for_layouts} is that the range for MITM layouts at the optimal separation is longer than for the RxTx layouts.
This is to be expected, since for the same separation between repeater nodes, the MITM layout performs strictly better than RxTx, thanks to its faster round trip times.
This optimal range is, however, achieved at rather low distances between network nodes (below 30~km for both codes in \cref{fig:range_for_layouts}).
We can see that, if we consider instead inter-node distances beyond 40~km, typical for long-distance fibre networks, the RxTx layout dominates.

Additionally, we notice that the larger Iceberg-2 code has a higher maximum range, but this maximum occurs at shorter inter-node separations when compared with the repetition code, in either layout.
This makes sense, since for shorter hops, the errors in the swapping circuit become more impactful with respect to the effects of memory decoherence, and these errors are largely mitigated by the error-detecting capabilities of the code.
That is, a stronger code may work best in denser networks, where nodes are closer together.

In contrast, the two-qubit repetition code outperforms the Iceberg-2 code for any node separation beyond 50~km.
This is because, at those distances, the long time it takes to generate a higher number of entangled pairs per hop greatly increases memory decoherence.
It is worth mentioning that the two-qubit code can distribute secret keys at a practical rate at a distance of 1000~km for inter-node separations up to slightly over 70~km; while the Iceberg-2 code can do the same when the node separation is below 60~km.
This suggests that, when nodes are spread apart, smaller codes could be the best option to enable long-distance quantum key distribution.

\subsection{Compensating for slow technological progress}
\label{sec:fixing_slow}

\begin{figure}
    \centering
    \includegraphics[width=0.99\linewidth]{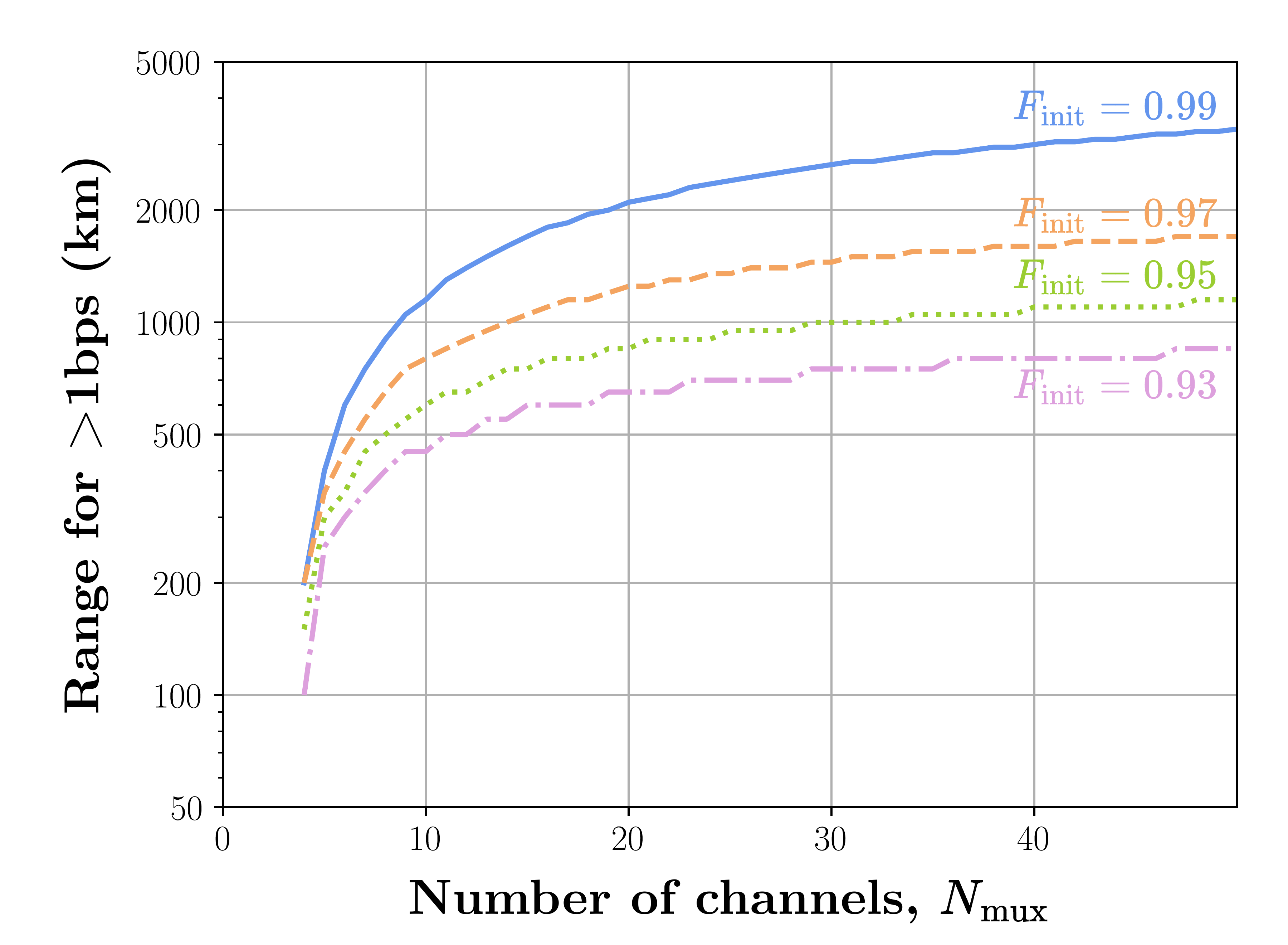}
    \caption{
    Range against the number of quantum channels, \(\numchannels\), for different values of \(\initfid\) when using the Iceberg-2 code.
    Memory decoherence is depolarizing, the BSM layout is RxTx and \(\hopdist = \SI{50}{\kilo\meter}\).
    }
    \label{fig:range_v_nchans}
\end{figure}

So far, we have investigated the design choices that optimize performance within some constrained settings.
Specifically, we have assumed that hardware with some desired parameters is available.
In this section, we look instead at the scenario in which one of these parameters is worse than its expected value in \cref{tab:parameters}, e.g. due to slow technological progress, and we improve a different parameter in an attempt to compensate it.

We focus on the particular case in which the initial fidelity, \(\initfid\), is lower than previously expected.
The reason is that improving this parameter in practical setups is considered particularly challenging.
Moreover, when \(\hopdist\) is in the order of tens of kilometres, techniques like link-level entanglement distillation~\cite{Bennett_Purification:1996,Deutsch_QPrivacy:1996} suffer from diminishing returns, where distributing additional entanglement to increase the fidelity results in more memory decoherence, thus counteracting the improvement.

In contrast, we shall assume that the number of quantum channels, \(\numchannels\), can be increased beyond 12.
Crucially, this parameter mainly impacts the waiting time and memory decoherence at each hop, much like others such as \(\probcou\) or \(\detefficiency\).
As a consequence, we use it here as a proxy for any improvements on the entanglement distribution procedure, and avoid discussing practical considerations concerning \(\numchannels\) in particular.

We plot in \cref{fig:range_v_nchans} the range at which a repeater chain can distribute more than 1 bit per second against \(\numchannels\) for different values of \(\initfid\), provided the Iceberg-2 code is used.
We see that, naturally, the practical range increases along with the number of channels, and decreases as the initial fidelity goes down.
More importantly, we see that key distribution can be achieved at practical rates for distances beyond 1000~km even for \(\initfid = 0.95\), provided the number of channels is increased to 30.
We consider this a relatively moderate increase, as we have tripled the resources to deal with an infidelity five times as large as the one in \cref{tab:parameters}.
For \(\initfid=0.93\), however, it appears that no amount of independent quantum channels is enough to extend the range up to 1000~km.
In fact, we generally observe in \cref{fig:range_v_nchans} that the effect of increasing \(\numchannels\) diminishes for smaller values of \(\initfid\), as the errors derived from memory decoherence become less dominant.

\begin{figure}
    \centering
    \includegraphics[width=0.99\linewidth]{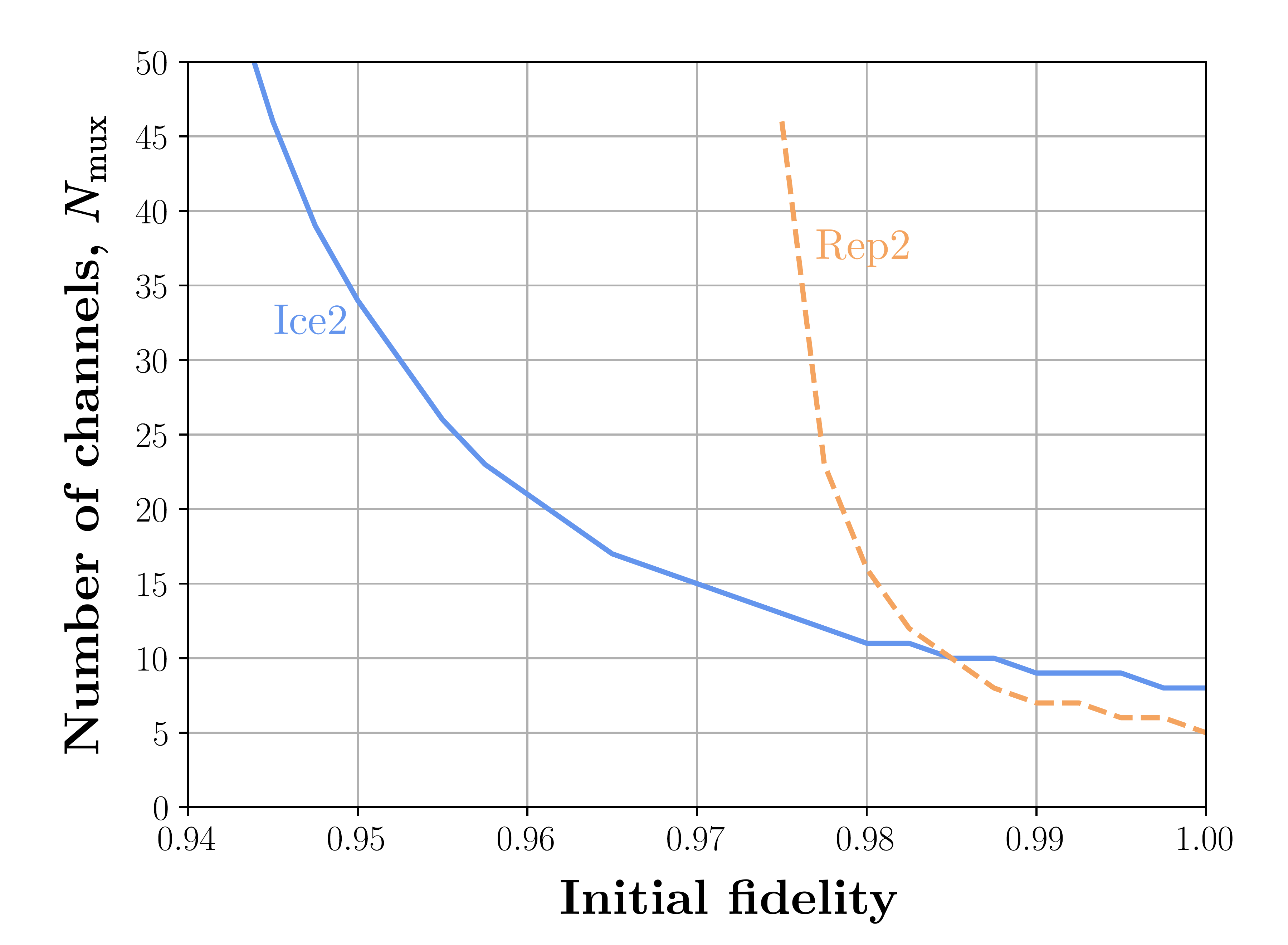}
    \caption{
    Required number of quantum channels to achieve \(\keyrate > \SI{1}{bps}\) at \(\totaldist = \SI{1000}{\kilo\meter}\) against the initial fidelity, for the two-qubit repetition (``Rep2") and Iceberg-2 (``Ice2") codes.
    Memory decoherence is depolarizing, the BSM layout is RxTx and \(\hopdist = \SI{50}{\kilo\meter}\).
    }
    \label{fig:required_nmux}
\end{figure}

This observation motivates us to investigate the relationship between these two quantities further.
Specifically, we want to check how many quantum channels are necessary to achieve a specific performance goal, given a value of \(\initfid\).
To do so, we plot in \cref{fig:required_nmux} the minimum number of channels \(\numchannels\) that enables practical quantum secret key distribution with \(\keyrate > \SI{1}{bps}\) at a distance of 1000~km, for both the Iceberg-2 and two-qubit repetition codes.
This figure shows that the number of channels required to achieve the set goal initially grows slowly as the infidelity becomes larger for both codes.
Then, as the errors in the initial entanglement errors become dominant over the effects of memory decoherence, the required number of channels quickly goes up.

\Cref{fig:required_nmux} shows that the required number of channels remains relatively low (\(\numchannels \leq 15\)) for fidelities above 97\%, provided the Iceberg-2 code is used.
In fact, this code can even enable practical key distribution at long distances when the initial fidelity is 94\%, given that \(\numchannels = 45\) independent channels are available.
In contrast, the number of required channels for practical QKD using the two-qubit repetition code is above 50 if \(\initfid = 0.97\), though the same requirement is relaxed to less than 12 if fidelities above 98\% can be guaranteed.
The reason for the apparent vulnerability of the two-qubit repetition code to infidelity is our assumption of the Werner profile in the initial entanglement.
Indeed, Werner states correspond to the ideal Bell pair undergoing a depolarizing channel, which may introduce bit-flips that are never detected, thus accumulating uncontrollably along the chain.
Notably, depending on the physical setup used to generate entanglement, the errors in the initial entanglement may come in different profiles which are more easily dealt with by the repetition code.

\section{Conclusion}
\label{sec:conclusion}

We have investigated key design choices for sequential entanglement generation with error detection (SEG-ED), using quantum key distribution over a repeater chain as a benchmark under near-term hardware assumptions.
To support this analysis, we have extended the framework introduced in \cite{ReyDominguez_Rethinking:2026} to enable the exact modelling of CSS-encoded repeater systems involving stabiliser circuits and Pauli-diagonal noise channels.
We have applied this framework to systems using repetition codes, the LNCY code~\cite{Leung_FourQubitCode1:1997} and the four-qubit code introduced in \cite{Vaidman_ErrorPreventionC422:1996, Grassl_CodesForErasureChannel:1997}, which we refer to as the Iceberg-2 code.

Our results highlight the remarkable performance of the two-qubit repetition code.
Despite being unable to detect bit-flip errors, this code substantially outperforms unencoded strategies and, under many of the near-term operating conditions considered here, also outperforms larger codes with more general error-detection capabilities.
This result is particularly relevant for early proof-of-principle demonstrations of encoded quantum repeaters, since implementing the two-qubit repetition code requires comparatively few physical qubits and relatively simple encoding, swapping, and decoding operations.
Its favourable performance therefore suggests that useful demonstrations of SEG-ED may not need to wait for the development of larger codes or more sophisticated codes.

The suitability of this simple code is further supported by our analysis of the entanglement-generation architecture.
When the available network sites are separated by approximately 50~km to 70~km, as may occur in typical long-distance fibre communications networks, integrating the optical Bell-state-measurement module into a memory-equipped repeater node is generally more favourable than deploying a separate module at the midpoint of each link.
Under these deployment constraints, the two-qubit repetition code remains particularly competitive.
The combination of a small code and an architecture that avoids additional intermediate devices could reduce both the hardware complexity and the infrastructure requirements of an initial SEG-ED implementation.

Nevertheless, the two-qubit repetition code is not optimal under all operating conditions.
In particular, its performance is more sensitive to imperfections in the initial entanglement when the associated noise is modelled as Werner-depolarizing, since such noise includes error components that the code cannot detect.
Instead, the LNCY or Iceberg-2 codes may become preferable in scenarios where the initial generation step is very noisy, as our analysis shows that these four-qubit codes can tolerate lower initial fidelities provided additional resources for generating entanglement in parallel are available.

Notably, the relative disadvantage of the repetition code may be less pronounced in physical platforms where the initial-entanglement noise is biased toward the type of error it detects.
Consequently, the choice between the two-qubit and four-qubit codes should depend not only on an aggregate fidelity value, but also on the underlying error profile.
This highlight the need for more in-depth platform specific investigations on this setup.

Overall, our results identify the two-qubit repetition code as a strong candidate for near-term SEG-ED implementations.
They also show that code selection cannot be separated from the noise model, entanglement-generation architecture, and constraints of the available telecommunications infrastructure.
Several effects remain to be incorporated into the analysis, including those related to multi-user traffic and resource sharing, queue buffering and previously mentioned platform-specific such as more detailed error models or restricted sets of operation.
Accounting for these effects will be necessary to determine whether the advantages identified here persist in complete repeater networks and to establish the complete design guidelines for experimental implementations.


\section*{Acknowledgments}\label{sec:acks}

We thank Nicolò Lo Piparo, Bill Munro, Kae Nemoto, Rodney van Meter and Ahmed Lawey for insightful discussions.
We acknowledge support from the European Union’s Horizon Europe Framework Programme under the Marie Sklodowska Curie Grant No. 101072637, Project Quantum-Safe Internet (QSI) and the UKRI EPSRC grant No. EP/X028313/1.

\section*{Data availability statement}\label{sec:data_avail}

All data that support the findings of this study are included within the article.

\appendix

\section{Diagonal representation of stabiliser states}
\label{app:diagonal_stabiliser}
{
\newcommand{\stabstate}{\phi_{\stabiliser}}

In this appendix, we explain the analytical framework used in the paper, previously introduced in \cref{sec:framework}.
We remark that, although we use this framework in the specific context of CSS-encoded repeaters, it can also be a valuable tool in the exact analysis of quantum setups composed of stabiliser circuits and Pauli-diagonal channels.

This appendix is structured as follows.
First, we give some background on the stabiliser formalism~\cite{Gottesman_HeisenbergRepresentation:1998,Aaronson_ImprovedSimulationOfStabilizer:2004} and introduce relevant definitions in Appendix \ref{app:stabiliser_background}.
Then, Appendix \ref{app:representation_sbasis} describes how to derive the diagonal representation of a stabiliser state, as well as how to simulate stabiliser circuits, Pauli-diagonal channels and measurements acting on it.
Lastly, we enumerate some common noise channels and express them as transition matrices in Appendix \ref{app:diagonal_error_models}. 

\subsection{Background}
\label{app:stabiliser_background}

Let us first define the basics of the stabiliser formalism~\cite{Gottesman_HeisenbergRepresentation:1998,Aaronson_ImprovedSimulationOfStabilizer:2004}.

\begin{definition}
    An \(m\)-qubit state, say \(\ket{\stabstate}\), is \emph{stabilised} by a Pauli operator \(\sigma \in \pgroup{m}\) if \(\sigma \ket{\stabstate} = \ket{\stabstate}\).

    The set \(\stabiliser\) of Pauli operators on \(m\) qubits which stabilise an \(m\)-qubit state is called the \emph{stabiliser group}, or simply stabiliser.
    The stabiliser constitutes an Abelian subgroup of \(\pgroup{m}\), \(\stabiliser \subset \pgroup{m}\), that does not contain the operator \(-\identity\).
\end{definition}

It is often useful to describe stabilisers, or other groups, using a generating set.
In particular, for \(\genset{G}\) any group with minimal generating set \(\{ g_1, \dots, g_r \}\), we may write \(\genset{G} = \braket{g_1, \dots, g_r}\).
We refer to the elements of the set as the \emph{generators} of \(\genset{G}\).

Consider now a stabiliser group \(\stabiliser = \braket{s_1, \dots, s_r}\) on \(m\) qubits, with \(r\) (non-unique) independent generators.
Then, the set of states stabilised by \(\stabiliser\), \(\genset{V}_{\stabiliser}\), constitutes a subspace \(\genset{V}_{\stabiliser} \subset \hilbert{m}\) of dimension \(2^{m-r}\), with \(\hilbert{m}\) denoting the Hilbert space on \(m\) qubits.
When \(r = m\), the stabiliser \(\stabiliser\) uniquely identifies a single state, up to a global phase.

\begin{definition}
    A stabiliser group on \(m\) qubits is called \emph{``full"} if it has \(m\) independent generators.

    An \(m\)-qubit state is called a \emph{stabiliser state} if its stabiliser group is full.
    That is, a stabiliser state is a state that can be uniquely determined by its stabiliser group.
\end{definition}

There is an important subclass of quantum operations which can be efficiently simulated when acting on stabiliser states.

\begin{definition}
    A unitary operation on \(m\) qubits, \(U\), is a \emph{Clifford operation} if it maps \(\pgroup{m}\) to itself under conjugation.
    That is, if \(U \sigma U^\adjoint \in \pgroup{m}\) for every \(\sigma \in \pgroup{m}\).

    A circuit consisting exclusively of preparations in the computational basis, Clifford operations and Pauli measurements is a \emph{stabiliser circuit}.
\end{definition}

The fact that stabiliser circuits can be efficiently simulated is a conclusion often called the Gottesman-Knill theorem~\cite{Gottesman_HeisenbergRepresentation:1998,Aaronson_ImprovedSimulationOfStabilizer:2004}.
Notably, the original proof of this theorem worked exclusively by keeping track of the generators of the state's stabiliser.
However, it was shown~\cite{Aaronson_ImprovedSimulationOfStabilizer:2004} that it is possible to improve the efficiency of this simulation by considering an additional list of generators, describing the so-called destabiliser group.

\begin{definition}
    Let \(\stabiliser = \braket{s_1, \dots, s_r}\) denote a stabiliser group on \(m\) qubits.
    A destabiliser group, \(\destabiliser \subset \pgroup{m}\), associated to \(\stabiliser\) is a stabiliser group on \(m\) qubits that can be generated as \(\destabiliser := \braket{d_1, \dots, d_r}\), such that \(d_i\) anti-commutes with \(s_i\) and commutes with all other generators of \(\stabiliser\), for all \(i = 1, \dots, r\).
    The destabiliser group associated to a stabiliser is not unique.
    
    If \(\stabiliser\) is a full stabiliser, we also refer to \(\destabiliser\) as full.
\end{definition}

Importantly, it is possible to define an orthonormal basis for \(\hilbert{m}\) with respect to any pair of full stabiliser and destabiliser groups~\cite{Yoder_GeneralizationOfTheStabilizer:2012}.

\begin{definition}
    Let \(\stabiliser = \braket{s_1, \dots, s_m}\) and \(\destabiliser = \braket{d_1, \dots, d_m}\) be a full stabiliser and an associated full destabiliser, respectively.
    Then, the \emph{stabiliser basis} of \(\hilbert{m}\) associated to the pair \(\{\stabiliser, \destabiliser\}\) is defined as \(\bbasis_{\stabiliser, \destabiliser} := \{ d \ket{\stabstate} |\ d \in \destabiliser \}\), where \(\ket{\stabstate}\) is the stabiliser state uniquely determined by \(\stabiliser\).
\end{definition}

Lastly, another property of full stabilisers and destabilisers is that they generate the Pauli group up to phase.
In fact, given a pair \(\{\stabiliser, \destabiliser\}\) of associated stabiliser-destabiliser groups on \(m\) qubits, it is possible to decompose any operator \(\sigma \in \pgroup{m}\) as \(\sigma = \alpha d s\) for some \(s \in \stabiliser\) and \(d \in \destabiliser\), where \(\alpha \in \{\pm 1,\pm i\}\).
We denote this by writing \(\sigma \in d\stabiliser\), where \(d\stabiliser\) is the coset of \(\stabiliser\) in \(\pgroup{m} / \{\pm 1, \pm i\}\) determined by \(d\).

\subsection{Representation in the stabiliser basis}
\label{app:representation_sbasis}

We now discuss the representation of quantum states as diagonal vectors in a stabiliser basis.
Furthermore, we show how to represent their transformations under Pauli-diagonal channels and Clifford operations, and how to simulate the probability distribution when measuring them.

Consider a stabiliser state on \(m\) qubits, \(\ket{\stabstate}\), associated to the stabiliser group \(\stabiliser = \braket{s_1, \dots, s_m}\).
Let \(\destabiliser = \braket{d_1, \dots, d_m}\) be a destabiliser group associated to \(\stabiliser\), and \(\bbasis_{\stabiliser, \destabiliser}\) the stabiliser basis associated to both groups.
Suppose the state is transformed under an arbitrary Pauli-diagonal channel, \(\qoperation\).
Then, the resulting state is diagonal in \(\bbasis_{\stabiliser, \destabiliser}\), and its density matrix can be written as:
\begin{equation}\begin{split}
\label{eq:rewrite_pauli_diagonal}
    \qoperation(\Ketbra*{\stabstate}) &=
    \sum_{\sigma \in \pgroup{m}}
    \lambda_\sigma \sigma \ketbra*{\stabstate} \sigma^\adjoint
    \\
    &=
    \sum_{d \in \destabiliser}
    \lambda'_d d \ketbra*{\stabstate} d^\adjoint ,
\end{split}\end{equation}
with \(\lambda_\sigma\) denoting the weight of the channel \(\qoperation\), as defined in \cref{eq:pauli_diagonal}, and \(\lambda'_d\) denoting the elements of the diagonal vector representing \(\rho\) in \(\bbasis_{\stabiliser, \destabiliser}\).
These weights are defined as:
\begin{equation}
\label{eq:weights_cosets}
    \lambda'_d =
    \sum_{\sigma \in d\stabiliser} \lambda_\sigma .
\end{equation}

Suppose now that, instead of being in a pure stabiliser state, the \(m\)-qubit system above was instead in an arbitrary mixture of states in \(\bbasis_{\stabiliser, \destabiliser}\), described by the diagonal vector \(\dvec\).
After undergoing the channel \(\qoperation\), the state remains diagonal in the same basis.
The reason is that applying a Pauli operator \(\sigma \in d\stabiliser\) to the state results in a permutation of the vector \(\dvec\).
This is because each coset \(d\stabiliser\) gets mapped, under multiplication by \(\sigma\), to a different coset, \(\sigma d \stabiliser\), and this mapping is the same for all operators in \(d\stabiliser\).
That is, if \(\sigma' \in d \stabiliser\), then \(\sigma\sigma' \in \sigma d \stabiliser\).
Therefore, it is possible to obtain the new diagonal vector, \(\dvec'\), as:
\begin{equation}
\label{eq:sum_permutations}
    \dvec' =
    \sum_{d \in \destabiliser}
    \lambda'_d \transmat_{\text{perm}(d)} \dvec ,
\end{equation}
where the weights \(\lambda'_d\) are defined as in \cref{eq:weights_cosets} and \(\transmat_{\text{perm}(d)}\) is the permutation matrix associated to \(d\).

Equivalently, the action of the complete Pauli-diagonal channel \(\qoperation\) can be described by the transition matrix \(\transmat_{\qoperation}\), defined as:
\begin{equation}
    \transmat_{\qoperation} =
    \sum_{d \in \destabiliser}
    \lambda'_d \transmat_{\text{perm}(d)} .
\end{equation}

Next, we explain the transformation of the diagonal state under a Clifford operation.
Consider the same initial state as in the last step, \(\dvec\), which is now subjected to a Clifford operation \(U\).
It turns out that this operation can be simulated efficiently, as for pure stabiliser states, and it results in a state with the same diagonal vector but expressed in a different stabiliser basis.
Specifically, the new stabiliser basis can be derived by conjugating the generators of the initial stabiliser and destabiliser groups by \(U\)~\cite{Yoder_GeneralizationOfTheStabilizer:2012}.

Finally, suppose all \(m\) qubits of the system whose state is described by \(\{\dvec, \bbasis_{\stabiliser, \destabiliser}\}\) are measured in the computational basis.
We now explain how to obtain the probability of an outcome \(z \in \{0, 1\}^m\), \(p_z\).
First, use standard tools from the stabiliser formalism to obtain the set, \(\genset{R}\), of possible outcomes when measuring all qubits of \(\ket{\stabstate}\) in the computational basis.
Note that, for any stabiliser state, there are \(2^{t}\) such outcomes, for \(t \leq m\), and these all occur with the same probability.
Then, the probability of outcome \(z\) can be obtained as:
\begin{equation}
    p_z =
    \frac{1}{2^t} \sum_{r \in \genset{R}}
    \lambda_{\paulix_z \paulix_r} ,
\end{equation}
where \(\paulix_z\) and \(\paulix_r\) are the Pauli-X operators associated with the respective outcomes \(z\) and \(r\) (e.g. for \(m=3\), ``000" is associated to \(\identity\), ``001" is associated to \(\identity\identity\paulix\), and so on) and \(\lambda_{\paulix_z \paulix_r}\) is the element in \(\dvec\) corresponding to the coset of \(\stabiliser\) in \(\pgroup{m} / \{\pm 1, \pm i\}\) to which \(\paulix_z \paulix_r\) belongs.
Equivalently, \(p_z = \vec{P}_z^\transpose \dvec\), where \(\vec{P}_z\) is the measurement vector associated to outcome \(z\), which contains the value \(2^{-t}\) in the positions of the elements of \(\dvec\) which contribute to the sum above, and zeroes in all other positions.
We remark that, although the procedure we have just explained is designed for measurements in the computational basis, any measurement which is equivalent to a Clifford operation followed by this measurement can be simulated by using the other tools described in this section to simulate the Clifford operation.

\subsection{Typical Pauli channels as transition matrices}
\label{app:diagonal_error_models}

Let us now illustrate the framework we have just introduced by discussing some common and very useful Pauli-diagonal channels.
These will be useful later to express the effects of the error models defined in \cref{sec:error_models}.

For any stabiliser basis \(\bbasis_{\stabiliser, \destabiliser}\), we define the transition matrices of the bit-flip and phase-flip channels with probability \(p\) on qubit \(\system{Q}\), respectively, as:
\begin{equation}\begin{split}
\label{eq:flip_transitions}
    \tmatbitflip{p}_{\system{Q}} &=
    (1 - p) \identity + p \tmatx_{\system{Q}} ,
    \\
    \tmatphaseflip{p}_{\system{Q}} &=
    (1 - p) \identity + p \tmatz_{\system{Q}} ,
\end{split}\end{equation}
where \(\identity\) is the identity matrix associated to \(\bbasis_{\stabiliser, \destabiliser}\), while \(\tmatx_{\system{Q}}\) and \(\tmatz_{\system{Q}}\) are the respective permutation matrices in \(\bbasis_{\stabiliser, \destabiliser}\) for the \(\paulix\) and \(\pauliz\) gates acting on qubit \(\system{Q}\).
Similarly, we define the transition matrix of the depolarization channel with probability \(p\) on qubit \(\system{Q}\) as:
\begin{equation}
\label{eq:depolarizing_transition}
    \tmatdepol{p}_{\system{Q}}
    \!=\!
    \left(1 - \frac{3p}{4} \right) \identity
    +
    \frac{p}{4} \left(
    \tmatx_{\system{Q}}
    + \tmatz_{\system{Q}}
    + \tmatx_{\system{Q}}\tmatz_{\system{Q}}
    \right) .
\end{equation}

}

\section{Encoded repeaters with CSS codes}
\label{app:analysis_appendix}

In this section, we specify the operations involved in each step of the repeater chain based on CSS codes, and provide a convenient description of these operations within the diagonal framework introduced in \cref{sec:framework}.
In particular, we first provide some common notation, and then discuss the encoding, swapping and decoding steps~\cite{Jiang_EncodedQR:2009, Jing_Simple:2021} in Appendices \ref{app:encoding}, \ref{app:swapping} and \ref{app:decoding}, respectively.

Besides the stabiliser basis \(\basisent\), which was already introduced in \cref{sec:main_derivation}, we define two other stabiliser basis, \(\basiszero\) and \(\basisplus\).
These correspond, respectively, to the stabiliser groups for states \(\ket{\logicalzero}\) and \(\ket{\logicalplus}\).

It is also useful to introduce here a shorthand to deal with both decoherence models at once.
More precisely, we note that the memory dephasing operation defined in \cref{eq:memory_dephasing} corresponds to a phase-flip channel with probability \([1 - \exp(-\tau/\coherencetime)]/2\), where \(\tau\) is the storage time.
Meanwhile, the memory depolarizing operation corresponds to its homonymous channel with probability \(1 - \exp(-\tau/\coherencetime)\).
These channels can be conveniently expressed as transition matrices based on the matrices from Appendix \ref{app:diagonal_error_models}.
In general, we denote the transition matrix describing the decoherence of a qubit \(\system{Q}\) stored for a duration \(\tau\) as \(\tmatdecoh{\tau}_{\system{Q}}\).

\subsection{Encoding}
\label{app:encoding}

\begin{figure}
    \centering
    \includegraphics[width=0.99\columnwidth]{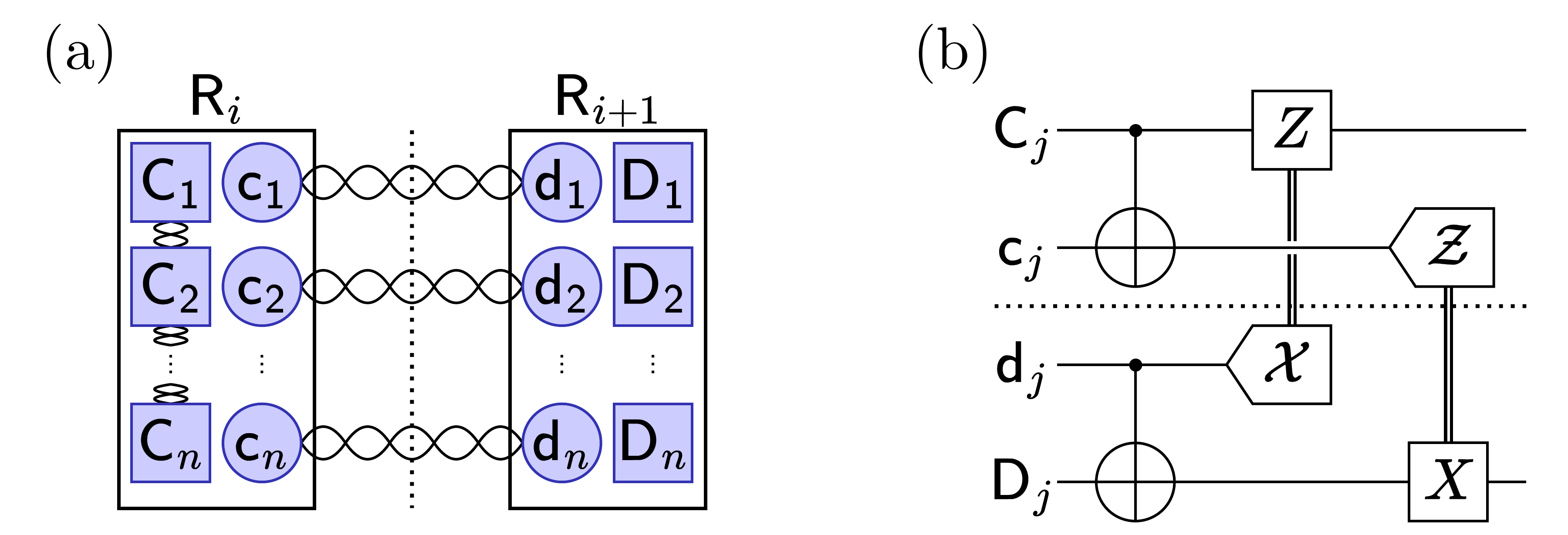}
    \caption{
    (a) Initial state before applying the encoding circuit between two repeater nodes.
    Here, squares (circles) denote data (communication) qubits and twisted lines denote entanglement.
    (b) Single row of the encoding circuit.
    }
    \label{fig:encoding_circuit}
\end{figure}

Consider two arbitrary nodes, \(\repeater{i}\) and \(\repeater{i+1}\), about to execute the encoding operation.
The data qubits in \(\repeater{i}\), labelled \(\system{C}_1\) through \(\system{C}_n\) in \cref{fig:encoding_circuit}(a), and those in \(\repeater{i+1}\), labelled \(\system{D}_1\) through \(\system{D}_n\), are prepared respectively in the diagonal states \(\{\shapeplus, \basisplus\}\) and  \(\{\shapezero, \basiszero\}\), where the values of the diagonal vectors \(\shapeplus\) and \(\shapezero\) are derived in Appendix \ref{app:data_preparation}.
Moreover, each pair of communication qubits is labelled \(\system{c}_1\)-\(\system{d}_1\) through \(\system{c}_n\)-\(\system{d}_n\) in order of the time at which they were successfully entangled.
The state of the \(j\)-th pair is diagonal in the Bell basis, which we denote by \(\bellbasis\), and we denote its diagonal vector by \(\shapepair{j}\).

The encoding step consists of employing the entanglement resources in the communication qubits to execute \(n\) remote CNOT gates, applied row-by-row as shown in \cref{fig:encoding_circuit}(b).
That is, the input to the operation consists of \(2n\) data qubits plus \(2n\) communication qubits.
Nevertheless, in order to minimise the dimension of the transition matrix representing the operation, we rearrange the elements of the operation so that we may first consider that \(n\) ideal CNOT gates are applied between the data qubits, followed by matrices describing the noise within \(\basisent\).
Specifically, we describe the operation in three steps:
(1) apply an ideal logical CNOT, which leaves the diagonal vector of the input state as is (i.e. \(\shapeplus \kron \shapezero\)) but updates the basis it is expressed in from \(\basisplus \otimes \basiszero\) to \(\basisent\);
(2) apply a series of transition matrices representing the effects of noise in each pair of communication qubits;
and (3) apply a transition matrix representing the noise originating from the circuit, i.e. during the measurements or the CNOT gates (see \cref{sec:error_models}).
These steps are summarised in \cref{eq:tmat_encoding}.

Let us first derive the values of \(\tmatencbell{j}\), for \(1 \leq j \leq n\), in \cref{eq:tmat_encoding}.
Each of these transition matrices can be derived from the diagonal vector of the state in each pair of communication qubits, \(\shapepair{j}\), and how the destabilisers in the stabiliser basis \(\bellbasis\) transform when they are rearranged after the ideal circuit.
In particular, suppose the diagonal vector for the \(j\)-th pair of communication qubits is \(\shapepair{j} = [ \lambda^{g_j}_\identity, \lambda^{g_j}_{\paulix_{\system{c}_j}}, \lambda^{g_j}_{\pauliz_{\system{d}_j}}, \lambda^{g_j}_{\paulix_{\system{c}_j}\pauliz_{\system{d}_j}} ]^\transpose\).
Then, the corresponding matrix describing the effects of its noise on the encoded state is:
\begin{equation}\begin{split}
\label{eq:tmat_noise_from_comms}
    \tmatencbell{j} &=
    \lambda^{g_j}_\identity \identity
    + \lambda^{g_j}_{\paulix_{\system{c}_j}} \tmatx_{\system{D}_j}
    + \lambda^{g_j}_{\pauliz_{\system{d}_j}} \tmatz_{\system{C}_j}
    \\
    &+ \lambda^{g_j}_{\paulix_{\system{c}_j}\pauliz_{\system{d}_j}} \tmatx_{\system{D}_j} \tmatz_{\system{C}_j} .
\end{split}\end{equation}

The diagonal vectors \(\shapepair{j}\) can be computed as:
\begin{equation}
\label{eq:shape_pair}
    \shapepair{j}
    =
    \tmatdecoh{\tgensuccess{n} - \tgensuccess{j}}_{\system{c}_j}
    \tmatdecoh{\tgensuccess{n} - \tgensuccess{j}}_{\system{d}_j}
    \begin{bmatrix}
        \initfid \\
        \frac{1 - \initfid}{3} \\
        \frac{1 - \initfid}{3} \\
        \frac{1 - \initfid}{3}
    \end{bmatrix} .
\end{equation}

Next, let us discuss circuit errors, that is, errors in the measurements and in the gates.
It can be shown that an error during the measurement of \(\system{c}_j\) induces a bit-flip in \(\system{D}_j\), due to the classically-controlled X gate.
Similarly, an error during the measurement of \(\system{d}_j\) induces a phase-flip in \(\system{C}_j\).
Therefore, the effects of these errors can be described using, respectively, the bit-flip and phase-flip channels with probability \(\probmeaserr\).
Errors in the CNOT gates, on the other hand, may be described as depolarizing channels with probability \(\probgateerr\).
Since all of these are Pauli-diagonal channels, we can rearrange them so that they are applied after the ideal circuit.
Doing this, we have that:
\begin{equation}\begin{split}
    \tmatnoisyenc &=
    \prod_{j=1}^n
    \tmatphaseflip{\probmeaserr}_{\system{C}_j}
    \tmatbitflip{\probmeaserr}_{\system{D}_j}
    \Big[
    (1 - \probgateerr)^2 \identity
    \\
    &+ (1 - \probgateerr) \probgateerr \tmatdepol{1}_{\system{C}_j} \tmatbitflip{1/2}_{\system{D}_j}
    \\
    &+ (1 - \probgateerr) \probgateerr \tmatphaseflip{1/2}_{\system{C}_j} \tmatdepol{1}_{\system{D}_j}
    \\
    &+ \probgateerr^2 \tmatdepol{1}_{\system{C}_j}\tmatdepol{1}_{\system{D}_j}
    \Big] .
\end{split}\end{equation}

\subsection{Swapping}
\label{app:swapping}
{

\newcommand{\destabphys}{\destabiliser_{\ket{\physicalbell{k}}}}
\newcommand{\destabarrange}{\widebar{\destabiliser}_{\ket{\physicalbell{k}}}}
\newcommand{\tmatout}[1]{\transmat^{\text{out}, {#1}}}
\newcommand{\tmatin}[1]{\transmat^{\text{in}, {#1}}}
\newcommand{\tmatadjust}[1]{\transmat^{\text{adj}, {#1}}}
\newcommand{\mvecbsm}[1]{\vec{P}^{\text{bsm} = {#1}}}
\newcommand{\mvecswpideal}{\vec{P}^{\text{SWP,ideal}}}
\newcommand{\tmatswperrs}{\transmat^{\text{SWP,err}}}

Let us focus now on the swapping operation in the virtual protocol.
Consider the same two repeater nodes as in Appendix \ref{app:encoding}, \(\repeater{i}\) and \(\repeater{i+1}\), whose data qubits store the state \(\{\shapeafterencoding, \basisent\}\), as described in that section.
Additionally, suppose \(\repeater{i}\) has a second set of data qubits, labelled \(\system{B}_1\) through \(\system{B}_n\) in \cref{fig:swapping_circuit}(a), which is logically entangled to \(n\) data qubits in Alice, labelled \(\system{A}_1\) through \(\system{A}_n\) in the picture.
Specifically, suppose that the state shared between Alice and \(\repeater{i}\) is \(\{\tmatmem\shapeafteriswaps{i-1}, \basisent\}\), where \(\shapeafteriswaps{i-1}\) is derived recursively via \cref{eq:shape_recursive}, and:
\begin{equation}
    \tmatmem =
    \prod_{j=1}^n
    \tmatdecoh{\thop}_{\system{B}_j} .
\end{equation}
For convenience, we define \(\shapeinputswpi{i} := \tmatmem\shapeafteriswaps{i-1} \kron \shapeafterencoding \) to denote the diagonal vector of the input to this stage, in the basis \(\basisent \otimes \basisent\).

\begin{figure}
    \centering
    \includegraphics[width=0.99\columnwidth]{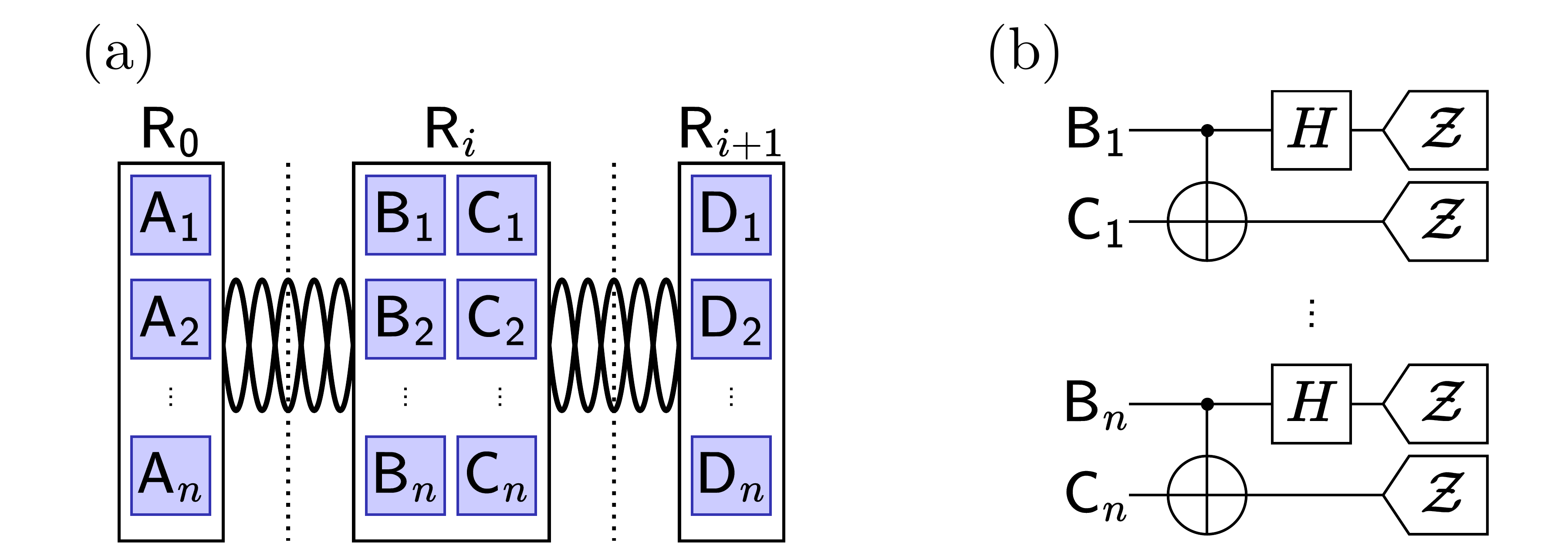}
    \caption{
    (a) Initial state before swapping two logical entangled states at some repeater node.
    Here, squares denote data qubits and thick twisted lines denote logical entanglement.
    (b) Circuit of the logical BSM, provided the logical \(\pauliz\) operators consist only of physical \(\pauliz\) operators.
    }
    \label{fig:swapping_circuit}
\end{figure}

In order to swap the two logical entangled states, a logical BSM is applied to the data qubits in \(\repeater{i}\), as shown in \cref{fig:swapping_circuit}(b).
Notably, the specific setup in \cref{fig:swapping_circuit}(b) is used when employing CSS codes whose logical \(\pauliz\) operators consist exclusively of physical \(\pauliz\) operators.
However, in some ``rotated" codes, such as the qubit-repetition code introduced in \cref{sec:code_comparison}, the logical \(\pauliz\) and \(\paulix\) operators swap places, with the former consisting only of physical \(\paulix\) operators.
In this case, the setup for the logical BSM can be simply adjusted by switching the basis in which we measure the qubits, that is, by applying a Hadamard rotation just before each measurement in  \cref{fig:swapping_circuit}(b).
Since we have assumed this can be applied without introducing errors (see \cref{sec:error_models}), we ignore this possibility and focus on the ``unrotated" setup. 

To avoid considering all \(4n\) qubits in the two logical entangled states while modelling the logical BSM, which acts exclusively on the \(2n\) data qubits within \(\repeater{i}\), we rearrange the input.
Firstly, let \(\destabphys\) denote the destabiliser of the physical state \(\ket{\physicalbell{k}}\), defined in \cref{eq:physical_bell}.
We define \(\destabarrange\) as the counterpart of \(\destabphys\) when acting on encoded states; for instance, if \(\destabphys\) includes an \(\paulix\) gate in the first physical qubit of \(\ket{\physicalbell{k}}\), then, \(\destabarrange\) includes a logical \(\paulix\) operator acting on the first logical qubit of \(\ket{\logicalbell}\).
Then, we have that:
\begin{equation}\begin{split}
\label{eq:rearrange_logical}
    &\Ket{\logicalbell}_{\system{AB}} \otimes \Ket{\logicalbell}_{\system{CD}}
    \\
    =
    &\frac{1}{\sqrt{|\destabarrange|}}
    \sum_{\sigma \in \destabarrange}
    \sigma_{\system{AD}} \Ket{\logicalbell}_{\system{AD}}
    \otimes \sigma^*_{\system{BC}} \Ket{\logicalbell}_{\system{BC}} ,
\end{split}\end{equation}
where we are using \(\system{A}:\equiv \system{A}_1\dots\system{A}_n\), \(\system{B}:\equiv \system{B}_1\dots\system{B}_n\), \(\system{C}:\equiv \system{C}_1\dots\system{C}_n\) and \(\system{D}:\equiv \system{D}_1\dots\system{D}_n\) for brevity.

Using \cref{eq:rearrange_logical}, and taking into account the Pauli-diagonal nature of the errors in the setup, we can prove that the state in \(\system{AD}\) after the BSM and the adjustments, provided the outcome is \(r\), is diagonal in \(\basisent\) with diagonal vector \(\shapeaftersres{i}{r}\), given by:
\begin{widetext}
\begin{equation}
    \shapeaftersres{i}{r} =
    \frac{1}{|\destabarrange|}
    \frac{
        \tmatadjust{r}
        \sum_{\sigma \in \destabarrange}
        \tmatout{\sigma}
        \diagonal\{
            \mvecbsm{r}
            \tmatin{\sigma}
        \}
        \shapeinputswpi{i}
    }
    { \probsres{r} } ,
\end{equation}
where
\(\tmatadjust{r}\) is the transition matrix denoting the adjustment conditioned on outcome \(r\);
\(\transmat^{\text{out}, \sigma}\) and \(\transmat^{\text{in}, \sigma}\) are matrices that map \(d_{\system{AD}}d_{\system{BC}} \ket{\logicalbell}_{\system{AB}} \ket{\logicalbell}_{\system{CD}}\) to \(d_{\system{AD}}\sigma_{\system{AD}} \ket{\logicalbell}_{\system{AD}}\) and \(d_{\system{BC}}\sigma^*_{\system{BC}} \ket{\logicalbell}_{\system{BC}}\), respectively, for every \(d_{\system{AD}}d_{\system{BC}}\) in the destabiliser of \(\ket{\logicalbell}^{\otimes 2}\);
\(\mvecbsm{r}\) is the measurement vector corresponding to the BSM outcome \(r\) when the input is diagonal in \(\basisent\);
\(\diagonal\{\vec{v}\}\) denotes a diagonal matrix with \(\vec{v}\) as its main diagonal;
and \(\probsres{r}\) is the probability that the outcome of the BSM is \(r\), which is given by:
\begin{equation}
\label{eq:prob_logbsm}
    \probsres{r} =
    \frac{1}{|\destabarrange|}
    \sum_{\sigma \in \destabarrange}
        \mvecbsm{r}
        \tmatin{\sigma}
        \shapeinputswpi{i} .
\end{equation}

\end{widetext}

Let \(2^{t_{x}}\) and \(2^{t_{z}}\) denote the number of outcomes identifying one particular codeword in the logical \(\xbasis\) and \(\zbasis\) bases, with \(t_x \leq n-k\) and \(t_z \leq n-k\).
There are \(|\destabarrange| = 4^k\) possible legitimate adjustments, and each of these is identified by any one of \(2^{t_x + t_z}\) equiprobable BSM outcomes, corresponding to a specific combination of codewords.
Notably, we can show that \(\shapeaftersres{i}{r_1} = \shapeaftersres{i}{r_2}\) and \(\probsres{r_1} = \probsres{r_2}\) for any two outcomes \(r_1, r_2\) of the \(2^{2k + t_x + t_z}\) corresponding to detecting no errors.
Therefore, it suffices to select one of those as the reference outcome to derive \(\shapeafteriswaps{i}\) and \(\probhop{i}\).
For convenience, we assume that this reference outcome is selected such that no adjustment is required and derive the remaining objects in \cref{eq:shape_recursive} as:
\begin{equation}
\label{eq:tmatswap}
    \tmatswp =
    \frac{1}{|\destabarrange|}
    \sum_{\sigma \in \destabarrange}
        \tmatout{\sigma}
        \diagonal\{
            \mvecbsm{\refresult}
            \tmatin{\sigma}
        \} ,
\end{equation}
and
\begin{equation}
\label{eq:mvecswap}
    \mvecswp =
    2^{t_x + t_z}
    \sum_{\sigma \in \destabarrange}
        \mvecbsm{\refresult}
        \tmatin{\sigma} ,
\end{equation}
where \(\refresult\) is the reference outcome.

Finally, we derive \(\mvecbsm{\refresult}\) as:
\begin{equation}
\label{eq:mvec_refswap}
    \mvecbsm{\refresult} =
    \mvecswpideal
    \tmatswperrs ,
\end{equation}
where \( \mvecswpideal \) is the measurement vector corresponding to outcome \(\refresult\) in the ideal setup,
and \( \tmatswperrs \) is the transition matrix representing the gate and measurement errors in the logical BSM circuit.
The value of \( \mvecswpideal \) can be derived using the techniques described in Appendix \ref{app:representation_sbasis}, while that of \( \tmatswperrs \) is given by:
\begin{equation}\begin{split}
\label{eq:swap_tmats}
    \tmatswperrs
    &=
    \prod_{j = 1}^n
    \tmatphaseflip{\probmeaserr}_{\system{B}_j}
    \tmatbitflip{\probmeaserr}_{\system{C}_j}
    \Big[
    (1 - \probgateerr) \identity
    \\
    &+ \probgateerr \tmatdepol{1}_{\system{B}_j} \tmatdepol{1}_{\system{C}_j}
    \Big] ,
\end{split}\end{equation}

} 

\subsection{Measurement-based decoding}
\label{app:decoding}
{
\newcommand{\mvecmeas}[3]{\vec{P}^{\text{meas}=({#2},{#3})|{#1}}}
\newcommand{\setdata}{\genset{S}_{\text{data}}}

In this section, we explain how to derive the vectors \(\mvecerate{\bbasis}{j}\), for \(j \in \{0, 1\}^k\) and \(\bbasis \in \{\zbasis, \xbasis\}\), defined in \cref{eq:error_rates}.
Firstly, we reiterate our previous point from Appendix \ref{app:swapping} regarding the measurements in ``rotated" codes, in which applying a Hadamard rotation before the measurement of each qubit is required.
Again, since this rotation can be applied without errors, we focus on the ``unrotated" case.
In the standard setup, errors in the measurements can be modelled as phase-flip or bit-flip channels with probability \(\probmeaserr\).
In particular, we compute:
\begin{equation}\begin{split}
\label{eq:mvec_erate}
    \mvecerate{\xbasis}{j}
    &=
    \!\!\!
    \sum_{\substack{
        \mathsf{k_a}, \mathsf{k_b} \in \{0,1\}^k \\
        \mathsf{k_a} \oplus \mathsf{k_b} = j
    }}
        \!\!\!\!\!
        \mvecmeas{\xbasis}{\mathsf{k_a}}{\mathsf{k_b}}
    \prod_{\system{Q} \in \setdata}
        \tmatphaseflip{\probmeaserr}_{\system{Q}} ,
    \\
    \mvecerate{\zbasis}{j}
    &=
    \!\!\!
    \sum_{\substack{
        \mathsf{k_a}, \mathsf{k_b} \in \{0,1\}^k \\
        \mathsf{k_a} \oplus \mathsf{k_b} = j
    }}
        \!\!\!\!\!
        \mvecmeas{\zbasis}{\mathsf{k_a}}{\mathsf{k_b}}
    \prod_{\system{Q} \in \setdata}
        \tmatbitflip{\probmeaserr}_{\system{Q}} ,
\end{split}\end{equation}
where \(\mvecmeas{\bbasis}{\mathsf{k_a}}{\mathsf{k_b}}\) is the measurement vector in \(\basisent\) corresponding to obtaining results \(\mathsf{k_a}\) and \(\mathsf{k_b}\) in Alice and Bob, respectively, provided they both measure in basis \(\bbasis\);
and \(\setdata\) is the set of the data qubits in Alice and Bob involved in the measurement.
The vectors \(\mvecmeas{\bbasis}{\mathsf{k_a}}{\mathsf{k_b}}\) can be derived using the methods laid out in Appendix \ref{app:representation_sbasis}.

} 

\section{Preparation of the data qubits}
\label{app:data_preparation}

\newcommand{\ghz}[1]{\text{GHZ}_{{#1}}}
\newcommand{\qubit}[1]{\system{Q}_{{#1}}}
\newcommand{\basisghz}[1]{\bbasis_{G{#1}}}
\newcommand{\basisz}[1]{\zbasis^{\otimes {#1}}}

\begin{figure}
    \centering
    \includegraphics[width=0.99\columnwidth]{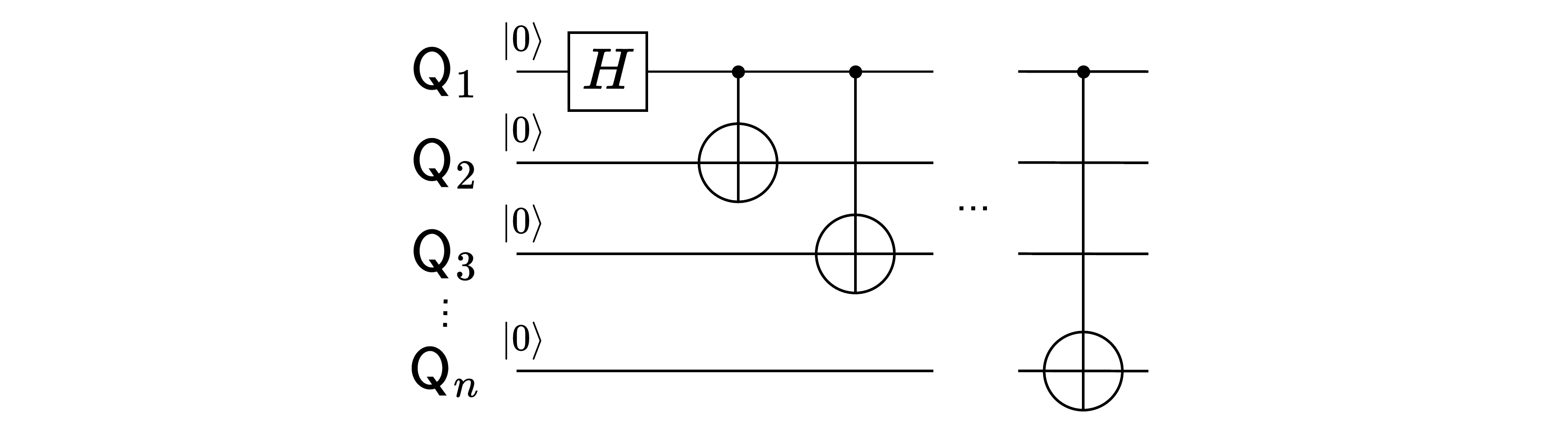}
    \caption{
    Preparation circuit for an \(n\)-qubit GHZ state.
    }
    \label{fig:ghz_preparation}
\end{figure}

In this section, we describe the circuits we have considered for the preparation of \(\ket{\logicalplus}\) and \(\ket{\logicalzero}\) in the codes introduced in \cref{sec:results}.
Remarkably, these states can be obtained by using a preparation based on the synthesis of an \(n\)-qubit GHZ state~\cite{Greenberger_GHZ:1989} for all codes.
The only exception is the logical zero state in the rotated repetition code, but this can be trivially prepared by initializing all data qubits to \(\ket{0}\) and then applying single-qubit Hadamard rotations.
To prove this, we first define the GHZ state and compare it against the states we aim to prepare in each code, and then detail the preparation circuit for the GHZ state.

The \(n\)-qubit GHZ state is a genuinely multipartite entangled state defined as
\begin{equation}
    \Ket{\ghz{n}} =
    \frac{1}{\sqrt{2}}
    \left( \ket{0}^{\otimes n} + \ket{1}^{\otimes n} \right) .
\end{equation}

By comparison, we see that \(\ket{\ghz{n}} = \ket{\phi_{0^n}}\) as defined in \cref{eq:bipartite_not_state}, where \(0^n\) denotes here an \(n\)-bit string of all zeroes.
This already reveals the relationship between the GHZ state and the codes considered in this paper.
Specifically, we have that the logical plus state in the rotated repetition code can be obtained by preparing \(\ket{\ghz{n}}\) and then rotating it with \(n\) Hadamard gates;
the logical plus state in the Iceberg-2 code is prepared in the same way, for \(n = 4\);
the logical zero state for the LNCY and the Iceberg-2 codes, which is identical, is simply \(\ket{\ghz{4}}\);
and, lastly, the logical plus state in the LNCY code corresponds to two Bell pairs in a product state, each of which is identical to \(\ket{\ghz{2}}\).

We next describe how to prepare this state.
While there exist some methods to generate local copies of this state using probabilistic local operations, speaking to the practicality of the codes considered here, we assume that we employ a deterministic gate-based method.
In particular, we consider the circuit in \cref{fig:ghz_preparation}, where we labelled the qubits \(\qubit{1}\) through \(\qubit{n}\) for convenience.
This circuits consists only of a single Hadamard gate, followed by \(n-1\) CNOT gates, so we are well-equipped to analyse it using the tools in Appendix \ref{app:diagonal_stabiliser}.

We analyse the circuit by rearranging the CNOT errors to the start of the circuit.
This way, before applying any ideal gate, we have the state \(\{\dvec_{\text{GHZ}, n}, \zbasis^{\otimes n}\}\), with diagonal vector:
\begin{equation}
    \dvec_{\text{GHZ}, n} =
    \prod_{j = 2}^n
    \left[
    (1 - \probgateerr) \identity
    + \probgateerr \tmatbitflip{1/2}_{\qubit{1}} \tmatbitflip{1/2}_{\qubit{j-1}} \tmatbitflip{1/2}_{\qubit{j}}
    \right]
    \dvec_{\mathds{1}} ,
\end{equation}
where \(\dvec_{\mathds{1}}\) denotes the trivial diagonal vector, with a single one as the first element.
To obtain the final state, it suffices to update the basis according to the rules in \cite{Yoder_GeneralizationOfTheStabilizer:2012}, while the diagonal vector remains unchanged.


\bibliography{references} 

\end{document}